# Predicting trajectories and mechanisms of antibiotic resistance evolution

Fernanda Pinheiro[1,♦], Omar Warsi[2,♦], Dan I. Andersson[2,]*, Michael Lässig[1,]*

[1] University of Cologne, Institute for Biological Physics, Köln, Germany
[2] Uppsala University, Dept. of Medical Biochemistry and Microbiology, Uppsala, Sweden
♦ Joint first authorship.
* Corresponding authors. Email: dan.andersson@imbim.uu.se, mlaessig@uni-koeln.de

**Bacteria evolve resistance to antibiotics by a multitude of mechanisms. A central, yet unsolved question is how resistance evolution affects cell growth at different drug levels. Here we develop a fitness model that predicts growth rates of common resistance mutants from their effects on cell metabolism. We map metabolic effects of resistance mutations in drug-free environments and under drug challenge; the resulting fitness trade-off defines a Pareto surface of resistance evolution. We predict evolutionary trajectories of dosage-dependent growth rates and resistance levels, as well as the prevalent resistance mechanism depending on drug and nutrient levels. These predictions are confirmed by empirical growth curves and genomic data of *E. coli* populations. Our results show that resistance evolution, by coupling major metabolic pathways, is strongly intertwined with systems biology and ecology of microbial populations.**

Modern human and veterinary medicine relies heavily on efficient antibiotics to prevent and cure bacterial infections in hospital and community settings. However, due to over- and misuse of antibiotics, we are faced with a global increase of antibiotic-resistant bacteria that reduces treatment efficacy and increases morbidity and mortality (*1*). Understanding how and under which conditions resistance evolves, and predicting likely trajectories of these dynamics are central goals in epidemiology, clinical microbiology, and evolutionary biology (*2-6*). Furthermore, predictive methods for resistance evolution are paramount for public health and drug developers to identify the least resistance-prone treatment protocols, drug targets, and new antibiotic candidates (*6*). Several recent advances put such predictions within reach. First, detailed experiments have established a number of genotype-fitness maps for antibiotic resistance and have mapped pervasive epistasis on these landscapes (*7-11*). Second, the evolutionary dynamics of resistance and other adaptive processes have been studied in parallel laboratory experiments (*8, 12-15*). These studies show that evolutionary change is often broadly distributed and heterogeneous at the genetic level; that is, the specific mutations occurring in one population are rarely seen in parallel-evolving populations. In contrast, functional targets and phenotypic effects of resistance are more repeatable, establishing an important prerequisite for predictability. Insight on the phenotypic basis of resistance comes from coupled growth, transcriptomics, and proteomics assays (*16-22*). These studies show that antibiotics cause a global perturbation of the cell's metabolic network and establish models relating core metabolic phenotypes to growth. Third, fitness models, which map phenotypic differences to selection, have been established as an important tool for evolutionary predictions (*4, 5*). To predict resistance evolution, we have to compute resistance and growth of mutants that establish under a given drug challenge, given input data of drug level and nutrient quality. To address this challenge, we develop metabolic fitness models for resistance evolution informed by parallel experiments in *E. coli* under different drug and nutrient levels. These models treat drug metabolism and evolutionary response as coupled perturbations of the cell's metabolic network. They successfully predict evolutionary trajectories of dosage-dependent resistance level and growth, including previously unseen resistance mutations.

**Resistance evolution experiments**

This study focuses on resistance evolution in response to the aminoglycoside streptomycin. Aminoglycosides act by binding to ribosomes and inhibiting translation. These broad-spectrum antibiotics are prescribed against different Gram-negative pathogens and mycobacterial infections (*23*). Streptomycin, in particular, has a well-studied mode of action, and quantitative models of its effects on cell metabolism and growth are available (*19*). Our selection experiments (Luria-Delbrück assays) elicit resistance mutants in the wild-type *E. coli* strain MG1655 at different drug and nutrient levels (rich medium, glycerol minimal medium). We apply drug concentrations moderately above the minimum inhibitory concentration on agar plates ($\text{MIC}^{\text{wt}} = 4$ mg/L), covering the range $0.9\times - 7.4\times$ half-inhibitory concentration in liquid culture ($d_{50}^{\text{wt}} = 8.7$ mg/L; this concentration reduces growth by 50%). Aminoglycosides require concentrations in this regime for clinical response and are typically toxic at higher levels (*23*). From each plated culture, we randomly pick eight different mutants for subsequent detailed analysis. For these mutants, we perform whole-genome sequencing and measure the growth rate $\lambda$ in liquid culture over the same range of drug levels ($d \leq 7.4\, d_{50}^{\text{wt}}$). Growth curves are recorded in units of the wild-type growth rate in a drug-free medium, $G(d) = \lambda(d)/\lambda_0^{\text{wt}}$.

Resistance mutations identified in this study are found to be broadly distributed over the *E. coli* genome, but they show striking regularities in their functional targets and growth patterns. In rich medium and at drug levels up to $\sim 5 \times d_{50}^{\text{wt}}$, most resistance mutations affect membrane-related functions associated, in particular, with the electron transport chain. Mutated genes and affected metabolic pathways are shown in Fig. S1 and listed in Table S1. Consistent with their common functional class, all membrane mutants have growth curves $G(d)$ with two similarities. First, there is a sizeable resistance cost, as measured by the drug-free growth rate relative to the wild type, $W = \lambda_0/\lambda_0^{\text{wt}}$. Second, drug response is marked by an initially slow decrease of growth followed by a rapid decline around the half-inhibitory concentration $d_{50}$ (Fig. S2, raw data are reported in Table S3). Importantly, this common pattern emerges despite considerable genomic variability; 82% of all membrane mutations are sequenced only in a single clone. Beyond $\sim 5 \times d_{50}^{\text{wt}}$, mutations of ribosome genes become most prevalent; these drug target mutants have amino acid substitutions in different residues of the S12 protein encoded by the *rpsL* gene.

**Drug metabolism and growth**

To explain the observed growth characteristics, we relate the empirical growth curves of membrane mutants to drug metabolism (*19*) (Fig. 1a). Cells maintain an intra-cellular drug level $d_{\text{int}}$ by active transport through the cell membrane (with rates $\gamma_{\text{in}}$ and $\gamma_{\text{out}}$) and growth-induced dilution (with rate $\lambda$). Intra-cellular drug molecules bind to ribosomes (with an equilibrium constant $K$), leaving only the unbound fraction of ribosomes, $p_u = 1/(1 + d_{\text{int}}/K)$, available for translation (*18, 19*). The resulting growth effect can be described in terms of Monod's law (*16*),

$$\lambda(\kappa_n, \kappa_t) = C \frac{\kappa_n \kappa_t}{\kappa_n + \kappa_t}, \tag{1}$$

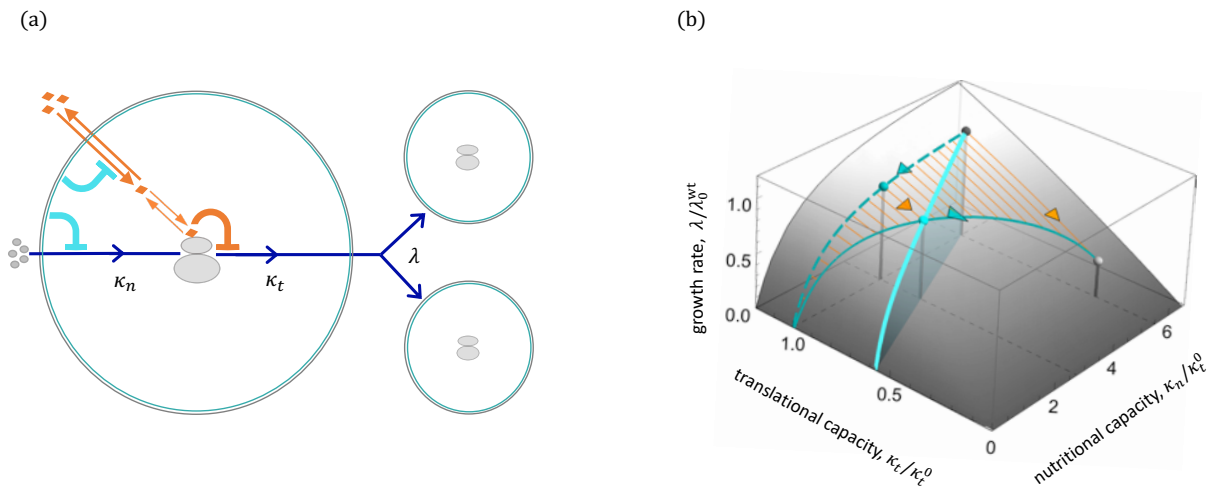

**Fig. 1. Drug metabolism and resistance evolution. (a)** Basic metabolism involves amino acid synthesis from extra-cellular nutrients and subsequent synthesis of proteins in ribosomes. These processes take place with rates $\kappa_n$ and $\kappa_t$ per unit of proteome, respectively, and result in cell growth with rate $\lambda$ (blue arrows). Ribosome-targeting drugs (here streptomycin) are transported through the cell membrane (with rates $\gamma_{\text{in}}$ and $\gamma_{\text{out}}$); intra-cellular drug molecules bind to ribosomes (with an equilibrium constant $K$) and impact growth by reducing $\kappa_t$ (*19*). Resistance evolution by membrane mutations (cyan) reduces the uptake rates of drug and nutrients, $\gamma_{\text{in}}$ and $\kappa_n$. **(b)** Monod's law describes the dependence of the growth rate on nutritional and translational capacity, $\lambda(\kappa_n, \kappa_t)$ (shaded surface). Drug effects (reduction of $\kappa_t$, orange arrows) and resistance evolution by membrane mutations (reduction of $\kappa_n$, cyan arrows) affect growth in mutually dependent ways. Our model establishes computable fitness landscapes for resistance mutants at a given drug level (cyan solid line) and in a drug-free medium (cyan dashed line). These landscapes predict a maximum-growth trajectory of resistance evolution as a function of drug concentration (light cyan line). Rates are shown relative to the wild type in a rich, drug-free medium (dark grey dot; $\kappa_t^0, \kappa_n^{\text{wt}} = 5.9\,\kappa_t^0, \lambda_0^{\text{wt}}$).

which relates the growth rate to the basic metabolism of the cell (Methods, Fig. 1). Here $\kappa_n$ is the nutritional capacity (the rate of synthesis of biomolecules from external nutrients), and $\kappa_t$ is the translational capacity (the rate of protein synthesis per ribosome); both rates are measured per unit of proteome. The pre-factor $C = \lambda_{\max}/\kappa_t$ is related to the maximum attainable growth rate. A ribosome-targeting drug impairs growth by reducing translation; that is, the growth rate follows Monod's law with constant $\kappa_n$ and drug-dependent $\kappa_t(d) = p_u(d)\,\kappa_t^0$, where $\kappa_t^0$ is the drug-free translational capacity (Fig. 1b). Specifically, the model of Greulich et al. (*19*) establishes computable drug response curves depending on two effective parameters: the rate scale $\lambda_* = 2(\gamma_{\text{out}}\,\kappa_t^0\,K)^{1/2}$, which determines the shape of the growth curve, and the concentration scale $d_* = \lambda_{\max}\lambda_*/(2\kappa_t^0\gamma_{\text{in}})$, which sets characteristic drug levels for growth inhibition. By fitting this model to growth data of membrane mutants, we obtain growth curves $G(d)$ that depend on the drug response parameters $d_*, \lambda_*$ and one model-independent fit parameter, the drug-free growth rate $W$ (Methods). Remarkably, the maximum-likelihood fits reproduce the growth data within experimental errors, validating the underlying metabolic model (*19*) for our membrane mutants (Fig. S2). For each mutant, the inferred parameters $d_*, \lambda_*$ yield estimates of the relative membrane transport rates, $\gamma_{\text{in}}/\gamma_{\text{in}}^{\text{wt}}$ and $\gamma_{\text{out}}/\gamma_{\text{out}}^{\text{wt}}$, and the fitted growth curve provides the empirical resistance level $R = d_{50}/d_{50}^{\text{wt}}$ (Table S2).

**Fitness model of drug resistance**

The mutant growth data reveals a simple, universal mechanism of membrane-based resistance evolution. This mechanism relates two a priori independent phenotypes: the resistance level, $R$, and the drug-free growth rate, $W$. We find both quantities to depend in a monotonic way on the relative drug uptake rate, $\varepsilon = \gamma_{\text{in}}/\gamma_{\text{in}}^{\text{wt}}$; this pattern is common to all membrane mutants except one outlier (Fig. 2ab). Moreover, the drug-free growth $W(\varepsilon)$ follows Monod's law at constant $\kappa_t$ (Fig. 2b, cf. Fig. 1b); hence, the parameter $\varepsilon$ also equals the relative nutritional capacity,

$$\varepsilon = \frac{\gamma_{\text{in}}}{\gamma_{\text{in}}^{\text{wt}}} = \frac{\kappa_n}{\kappa_n^{\text{wt}}} \ . \tag{2}$$

Together, membrane resistance mutations generate a proportional reduction of drug and nutrient uptake depending on a single permeability parameter $\varepsilon$. This coupled effect is consistent with the underlying molecular transport processes; nutrients present in rich growth media and aminoglycosides both require active uptake mechanisms involving proton motive force (*23*). Combined with the model of drug metabolism (*19*), equation (2) specifies the growth rate or absolute fitness, $G(d, \varepsilon)$, of mutants with permeability $\varepsilon$ at drug level $d$ (Methods). This fitness model predicts the drug-free growth rate and resistance level for mutants of effect $\varepsilon$,

$$W(\varepsilon) = \frac{\varepsilon(q^{\text{wt}} + 1)}{(\varepsilon q^{\text{wt}} + 1)}, \qquad R(\varepsilon) = \frac{r^{\text{wt}}W(\varepsilon) + 1/(r^{\text{wt}}W(\varepsilon))}{\varepsilon(r^{\text{wt}} + 1/r^{\text{wt}})}, \tag{3}$$

depending on the wild type parameters $q^{\text{wt}} = \kappa_n^{\text{wt}}/\kappa_t^{\text{wt}}$ and $r^{\text{wt}} = \lambda_0^{\text{wt}}/\lambda_*^{\text{wt}}$ (Fig. 2ab). With values $q^{\text{wt}} = 5.9 \ (5.5, 6.3)$, $r^{\text{wt}} = 5.4 \ (5.3, 5.6)$ inferred from growth data, the relations (3) generate a unique evolutionary tradeoff $W(R)$, which relates the resistance effect of membrane mutations and the collateral cost in drug-free growth (Fig. 2c). Without further fit parameters, the model predictions are in quantitative agreement with the empirical data reported in Fig. 2. Importantly, the tradeoff $W(R)$ depends on the underlying evolutionary mechanism of equation (2) and on broad metabolic characteristics of the wild type; however, it does not involve genetic details of specific resistance mutations. In particular, the initial slope $W'(R = 1) = -(q^{\text{wt}} + 2/(1 + (r^{\text{wt}})^2))^{-1}$, predicts a universal weak-effect tradeoff: starting from our susceptible wild-type *E. coli* strain, 10% resistance increase costs 2% in drug-free growth.

To further characterize membrane-based resistance evolution, we assess the effects of drug release by outward transport. The inferred variation in $\gamma_{\text{out}}$ across our set of mutants turns out to have negligible effects on growth (Fig. S3), as confirmed by model variants that include evolution of $\gamma_{\text{out}}$ (Fig. S5). This weak dependence reflects the metabolic regime relevant for our experiments, where intra-cellular drug levels are depleted predominantly by growth and not by outward transport (*19*). It is also consistent with the observed molecular targets of membrane mutations. These affect membrane functions associated with electron transport but do not overlap with known efflux pathways (Fig. S1).

**Predicting trajectories of resistance evolution**

Next, we use the fitness model to predict resistance evolution in response to a given drug challenge. At a given drug level $d$, the growth rates $G(d, \varepsilon)$ define a fitness landscape for membrane mutants of different permeability (i.e. uptake rate) (Fig. 1b). There is a specific drug level $d_c(\varepsilon)$ where mutants of permeability $\varepsilon$ have maximum fitness (i.e., relative growth rate) compared to all other

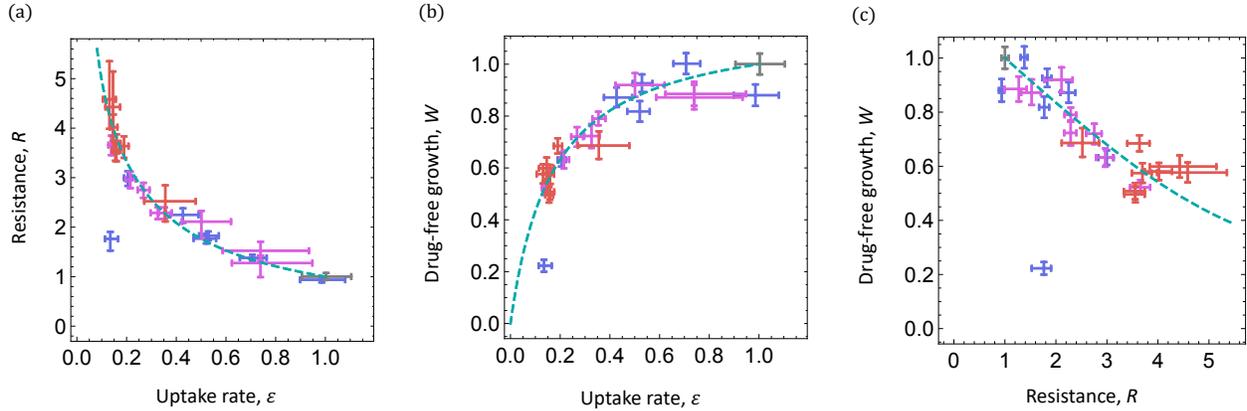

**Fig. 2. Resistance evolution by membrane mutations. (a, b)** Resistance, $R = d_{50}/d_{50}^{wt}$, and drug-free growth, $W = \lambda_0/\lambda_0^{wt}$, are plotted against the evolutionary change in membrane uptake rates, $\varepsilon = \gamma_{in}/\gamma_{in}^{wt} = \kappa_n/\kappa_n^{wt}$; see equation (2). **(c)** Evolutionary resistance-cost trade-off, obtained by plotting $W$ against $R$. Data points are obtained from experimental growth curves of membrane mutants (Methods, Table S2); bars give rms. measurement errors. Mutants are elicited in Luria-Delbrück assays at drug levels $d_{LD}/d_{50}^{wt} = 0.9$, 1.8, 3.7 (violet, pink, red); the wild type is shown for reference (gray). Model predictions (dashed cyan lines) are given by equation (3); the predicted drug-free growth rate $W(\kappa_n/\kappa_n^{wt})$ follows Monod's law at constant $\kappa_t$ (cf. Fig. 1b). Growth condition: rich LB, $\lambda_0^{wt} = 2.0/hr$, $d_{50}^{wt} = 8.7mg/L$. Model parameters: $q^{wt} = 5.9$, $r^{wt} = 5.4$.

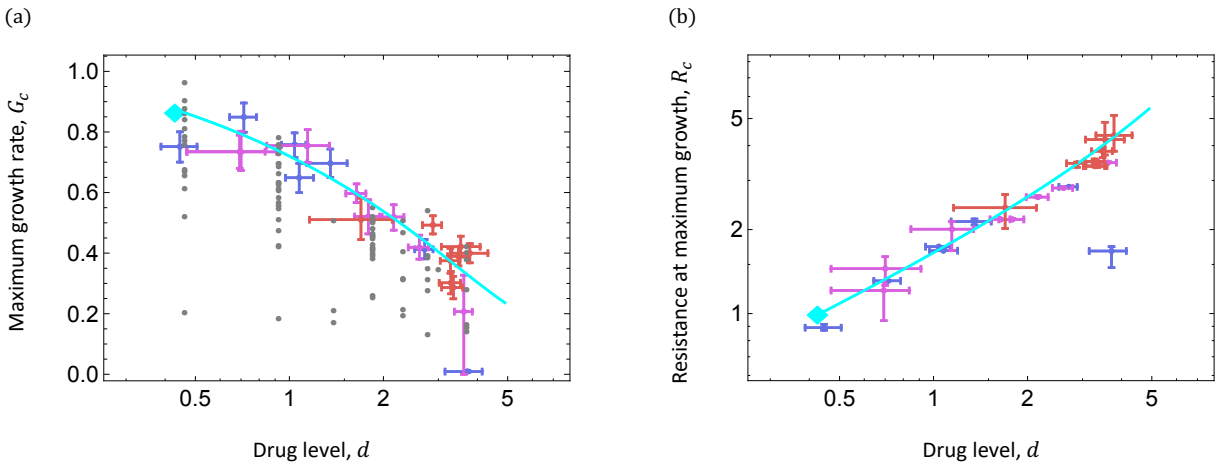

**Fig. 3. Predicting mutant growth and resistance. (a)** The predicted maximum-growth trajectory, $G_c(d)$ (light cyan line, cf. Fig. 1b) with onset point for resistance evolution (diamond) is compared to data of membrane mutants in liquid culture. For each mutant, empirical $G$ and $R$ values are shown at the predicted maximum-fitness point (dots, colors as in Fig. 2, bars give rms. experimental errors). Almost all other growth measureents (gray points) are below the predicted maximum-growth trajectory. **(b)** The predicted resistance of maximum-growth mutants, $R_c(d)$, is compared to empirical $R$ values shown at the predicted maximum-fitness point. Growth conditions: as in Fig. 2.

membrane mutants. We can express this growth rate,

$$G_c(\varepsilon) = \frac{\varepsilon \, (q^{\text{wt}} + 1)}{\varepsilon q^{\text{wt}} + 2}, \tag{4}$$

and the corresponding drug level $d_c(\varepsilon)$, in terms of the wild type parameters $q^{\text{wt}}, r^{\text{wt}}$ (Methods). Plotting $G_c(\varepsilon)$ vs. $d_c(\varepsilon)$ produces a key model prediction: the maximum growth trajectory of membrane resistance mutants at a given drug challenge, $G_c(d)$ (Fig. 3a, see also Fig. 1b). Geometrically, this trajectory is obtained as the envelope of the underlying family of growth curves. Together with equation (3), we can also compute the resistance of maximum-fitness mutants, $R_c(d)$ (Fig. 3b). To test these predictions, we plot the empirical growth rate of each membrane mutant at its predicted maximum-fitness drug level, $G(d_c)$ (Fig. 3a), and we plot the empirical resistance $R$ against $d_c$ (Fig. 3b). Both empirical patterns are in quantitative agreement with model predictions.

When will these maximum-fitness mutants be seen in selection and evolution experiments? To address this question, we evaluate the rate of resistance evolution in Luria-Delbrück assays (Fig. 4a). At moderate drug levels, we find resistance origination rates $U_m(d) = 10^{-7} - 10^{-9}$, from which we infer membrane mutation rates that decrease with increasing effect $R$ (Methods, Fig. S5). Together with our fitness model, this mutational spectrum predicts the range of growth rates likely to be observed in Luria-Delbrück selection experiments (shown as gray squares in Fig. 4b). These rates cluster into a narrow corridor below the maximum growth rate $G_c(d)$, which is constrained by two effects: mutations of larger resistance effects rapidly become rare, whereas mutations of smaller effects have rapidly decreasing growth rates (Fig. S5). Observed growth rates consistently cluster into this corridor; a similar clustering around $R_c(d)$ is observed for resistance effects (Fig. S5). Under subsequent short-term evolution at the same drug level, clonal interference between resistant clones is expected to further prune growth and resistance spectra towards the optimal values $G_c(d)$ and $R_c(d)$.

**Evolutionary switches between resistance mechanisms**

Beyond growth and resistance predictions for a given mechanism, evolutionary models can also predict the prevalent resistance mechanism itself. First, the wild type increases its ribosome content in response to ribosome-targeting drugs (*17, 19*), $\rho^{\text{wt}}(d) \sim \kappa_n^{\text{wt}}/(\kappa_t^{\text{wt}}(d) + \kappa_n^{\text{wt}})$, which can be regarded as a resistance mechanism based on regulation. This mechanism confers a substantial growth benefit at low drug levels, as can be seen by comparing the wild type growth curve with its counterpart in a hypothetical cell with a constant, unregulated ribosome content (Fig. 4a). Resistance by regulation is the dominant mechanism at low drug levels, because any evolutionary change with a cost in a drug-free environment ($W < 1$) can only be under positive selection beyond a minimum drug level (*3*).

Our membrane resistance model predicts an evolutionary switch from regulation to membrane evolution with an onset point $d_{\text{rm}}(q^{\text{wt}}, r^{\text{wt}}) = d_c(\varepsilon = 1)$ (Fig. 4b). At our wild type parameters, the predicted onset point $d_{\text{rm}}/d_{50}^{\text{wt}} \sim 1/2$ is consistent with the experimentally observed onset of membrane-based resistance evolution below the wild-type MIC (*9*). At higher drug levels, $d/d_{50}^{\text{wt}} > 5$, membrane mutations have low mutation rates ($U_m(d) < 10^{-10}$, Fig. S5) and low model-based growth rates $G_c(d)$ (below the detection threshold in a Luria-Delbrück assay for $d/d_{50}^{wt} > 10$), predicting another switch to a viable resistance mechanism. That switch is observed: in the regime $d/d_{50}^{wt} = 5 - 12$, most resistant clones have mutations in the *rpsL* gene, which encodes the ribosomal protein S12 (Fig. 4a). These target mutations confer high resistance with a flat maximum growth rate $G_t = 0.95$ in this drug regime, albeit at low mutation rate, $U_t \sim 10^{-10}$.

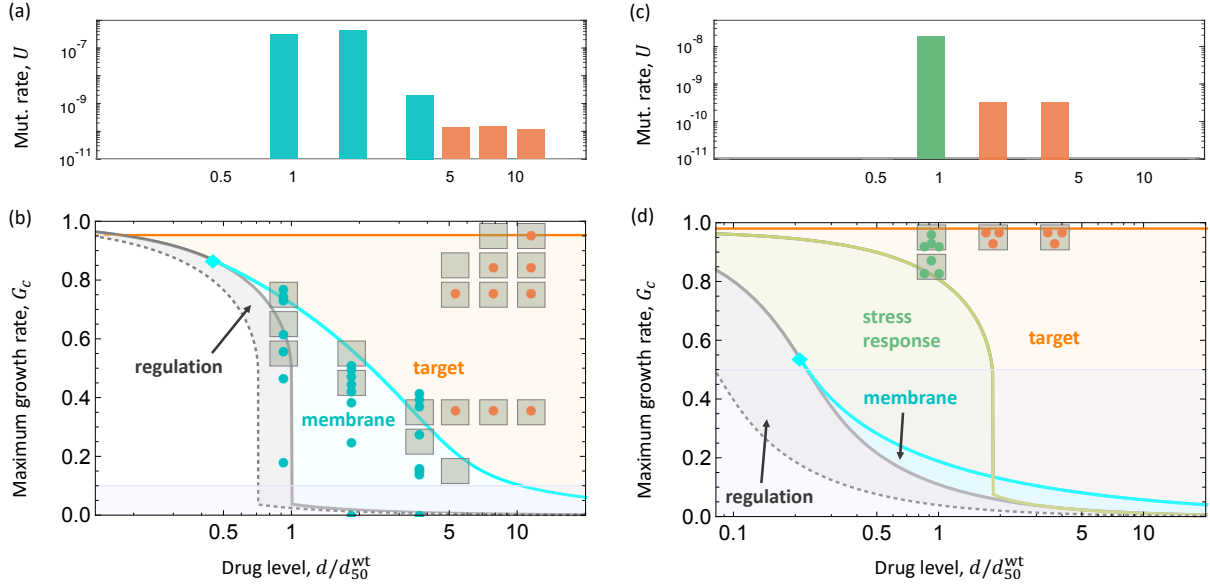

**Fig. 4. Predicting mechanisms of drug resistance. (a, c)** Rates and phenotypes of resistance mutations in rich agar and minimal glycerol at different drug levels (cyan: membrane mutations; orange: *rpsL*, ribosome target mutations; green: *cpxA*, stress response). **(b, d)** Prevalent growth rates and resistance mechanisms depend on drug and nutrient levels. Observed growth rates (measured in liquid culture) of Luria-Delbrück mutant colonies are shown together with the predicted corridor of likely growth rates (gray squares show growth rate intervals with probability ≥ 0.04, see also Fig. S5). Model predictions of maximum-growth trajectories for different resistance mechanisms: regulation of ribosome content in the wild-type (gray line) compared to hypothetical cell with constant ribosome content (dotted gray line), membrane permeability reduction (light cyan line) with onset point $d_{rm}$ in rich medium (diamond), *cpR* stress response system (green line, fitted growth curve, see Fig. S2). The growth rate ranking of mechanisms is emphasized by color shading. Horizontal lines mark a low-growth regime where different metabolic regimes are expected. Growth conditions and model parameters: rich medium ($\lambda_0^{wt} = 2.0$/hr, $d_{50}^{wt} = 8.7$mg/L used as unit for all drug levels, $q^{wt} = 5.9$, $r^{wt} = 5.4$), glycerol minimal medium ($\lambda_0^{wt} = 0.39$/hr, $q^{wt} = 0.2$, $r^{wt} = 1.03$).

Our model predicts a similar grading between resistance mechanisms in short-term evolution (Fig. S5). At intermediate drug levels and population sizes ($N \gtrsim U_m^{-1}(d) \sim 10^7 - 10^9$), resistant clones are expected to harbor membrane mutations but no target mutations, implying growth rates of order $G_c(d)$ as discussed above. This regime is relevant for infections; for example, typical urinary tract and blood infections involve populations of size $N < 10^9$ (24, 25). In very large populations ($N \gtrsim U_t^{-1}$), membrane mutations will be outcompeted by target mutations with growth rate $G_t$. Importantly, metabolic fitness models predict the optimal resistance mechanism to depend not only on the drug level, but also on nutrient conditions. Specifically, membrane mutations reduce $\kappa_n$ by affecting nutrient intake, which is an affordable resistance mechanism in rich media but becomes costly in poor media. Conversely, ribosome target mutations reducing $\kappa_t$ have higher metabolic costs in rich media (Fig. 1b), in accordance with previous results (*26*). This predicts a shift in prevalence towards target mutations with decreasing nutrient quality. To test this prediction, we select resistance mutants by Luria-Delbrück assays in a M9-minimal glycerol medium, which has a 5-fold reduced wild-type growth rate compared to rich medium (Fig. 4cd). The resulting model-based maximum growth rates $G_c(d)$ of membrane permeability mutations are below the detection threshold for all drug levels. Consistently, these mutations are never observed despite their substantial mutation rate, and target mutations become the prevalent resistance mechanism already at intermediate drug levels. At low drug levels, target mutations are superseded by a new class of low-cost, intermediate-rate ($U_{cpx} \sim 10^{-8}$) resistance mutations in the gene *cpxA*. These mutations activate the *Cpx* stress response system (*27*), which confers resistance to aminoglycosides (*28*) and affects nutrient intake by regulating porins (*29*). Because drug and nutrients are affected via different pathways, this resistance mechanism can lead to higher growth rates than a proportional-reduction mechanism constrained by equation (2).

More broadly, metabolic fitness models can serve to rank different resistance mechanisms. For example, overexpression of efflux pumps is a ubiquitous resistance mechanism observed in response to many antibiotics, including aminoglycosides, fluoroquinolones, tetracyclines and $\beta$-lactams (*30, 31*). Here we use a minimal fitness model with coupled effects on efflux rates and on proteome allocation to estimate maximum-growth trajectories for this mechanism (Fig. S4). For the growth conditions of our system, these trajectories remain below those of membrane-based resistance throughout the intermediate drug regime. The predicted inefficiency of efflux pumps is a consequence of the underlying metabolic regime, where intracellular drug is depleted predominantly by growth (*19*). It is consistent with the absence of efflux mutants in our Luria-Delbrück selection experiments. Notably, shape and parameter dependence of growth trajectories differ considerably between mechanisms, which opens the perspective of using empirical growth curves together with metabolic modeling to infer properties of unknown resistance pathways.

**Conclusion**

We have established an evolutionary model of drug resistance that predicts tradeoff functions $W(R)$ and maximum-growth trajectories $G_c(d)$ for families of mutants carrying specific resistance mechanisms. Tradeoff functions can be regarded as Pareto fronts constraining the simultaneous optimization of drug resistance and growth in a drug-free medium (*32, 33*). Maximum growth trajectories determine optimal resistance levels for a given mechanism and predict evolutionary switches between mechanisms as a function of drug and nutrient conditions. Such Pareto fronts and maximum-growth trajectories apply to short-term evolution by common resistance mutants. This regime is critical for the immediate survival of a microbial population under a drug challenge and relevant in clinical contexts. However, rare mutants operating on different mechanisms can reach over the Pareto front, and over longer evolutionary times, compensatory mutations are expected to mitigate resistance costs. Thus, including multiple mutations and compensatory traits into metabolic fitness models of resistance is an important avenue for future research (*9, 11*).

Predictive analysis of resistance evolution can help drug and therapy design by identifying likely resistance pathways for detailed experimental scrutiny. Our method is applicable to a wide range of antibiotics and resistance mechanisms, depending on two key prerequisites of predictability.

First, the dominant resistance trait has a sufficient mutational supply covering a spectrum of resistance effects, so that high-fitness resistance mutants are continuously present in a typical population. This condition is fulfilled in our experiments, as well as for a number of clinically relevant antibiotics, including fluoroquinolones and aminoglycosides. Predictability of resistance levels and growth rates is then generated by functional convergence towards an efficient resistance mechanism and tuning of this mechanism on its Pareto front (Fig. 3 and 4). Thus, predictability extends to previously unseen mutations, in tune with the broad genomic distribution of adaptive mutations observed here and in other studies (*5, 14*). Second, we have a quantitative biophysical model for drug action and resistance mechanisms. The arguably best-understood case of ribosome-targeting drugs (*18, 19*) is the focus of this paper; similar models appear within reach for other drugs targeting core metabolic pathways, such as the fluoroquinolone ciprofloxacin (*34*), as well as for drug combinations (*35*).

Our results highlight that drug action and resistance evolution are global, mutually dependent perturbations of a microbial system (Fig. 1b). Both act on the cell's metabolic network, but often on different components. The central role of cell metabolism for resistance evolution explains a key finding of this paper: optimal resistance effects and mechanisms strongly depend on the ecology of the microbial population, in particular, on its nutrient supply. Accounting for such environmental effects is crucial for translating laboratory experiments on resistance into clinical applications.

32. O. Shoval *et al.*, Evolutionary trade-offs, Pareto optimality, and the geometry of phenotype space. *Science* **336**, 1157-1160 (2012).
33. Y. Li, D. A. Petrov, G. Sherlock, Single nucleotide mapping of trait space reveals Pareto fronts that constrain adaptation. *Nature Ecology & Evolution*, 1-13 (2019).
34. N. Ojkic *et al.*, A roadblock-and-kill model explains the dynamical response to the DNA-targeting antibiotic ciprofloxacin. *BioRxiv*, 791145 (2020).
35. B. Kavčič, G. Tkačik, T. Bollenbach, Minimal biophysical model of combined antibiotic action. *bioRxiv*, (2020).
**Acknowledgements**.
We acknowledge discussions with Tobias Bollenbach, Arjan de Visser, Daniele Marmiroli and Isabel Gordo.

**Funding**.
This work has been partially funded by Deutsche Forschungsgemeinschaft grant CRC 1310 (to ML) and Swedish Research Council grant 2017-01527 (to DIA).

**Author Contributions.**
Funding acquisition: ML, DIA; Investigation: all authors; Writing – original draft: all authors; Writing – review & editing: all authors.

**Competing Interests.**
The authors declare no competing interest.

**Data and materials availability.** Correspondence and requests for materials should be addressed to mlaessig@uni-koeln.de or dan.andersson@imbim.uu.se.
11

## Methods

1. **Experimental procedures**

**Bacterial strains, growth media**. The bacterial strain used in these experiments was *Escherichia coli* K-12 MG1655. The streptomycin-resistant clones derived from this ancestral strain are listed in Table S3. Experiments in rich nutrient medium were performed in LB (liquid medium) and LB agar (solid medium, Sigma-Aldrich), and experiments in poor nutrient medium were performed in 0.2% glycerol M9-minimal medium (*36*). All experiments were performed at 37 °C. Minimum inhibitory concentrations (MIC) under different nutrient conditions were determined by plating a $10^{-7}$ dilution of an overnight-grown culture at different concentrations of streptomycin. For the wild-type strain, we measured a streptomycin MIC of 4 mg/L in rich agar and 2 mg/L in M9-minimal glycerol agar.

**Mutation rate data and isolation of mutants.** We determined the mutation rate of resistance at different streptomycin concentrations from classical Luria-Delbrück fluctuation assays. At intermediate drug concentrations used for selection in LB agar ($d = 8, 16, 32, 64$ mg/L), we started 40 independent cultures of $10^3$ cells from an overnight culture. From the number of colonies and the mean number of plated cells across these 40 parallel cultures, we computed drug-dependent mutation rates, using the bz-calculator for mutation rates (*37*). Colonies obtained from these tests were re-streaked on plates with the same streptomycin concentration. To determine the mutation rates of *rpsL* variants, we plated 20 – 50 independent overnight cultures of $10^3$ initial cells at higher drug levels ($d = 48, 64, 100, 128$ mg/L). In minimal glycerol medium, we used 10 – 20 independent cultures at $d = 8, 16, 32$ mg/L. Mutation rates and frequencies of different variants are given in Table S5.

**Growth rate measurements.** Measurements in the exponential growth phase were performed with a Bioscreen plate reader at $OD_{600}$, with measurements taken every 4 minutes. To calculate the maximum exponential growth rate from the $OD_{600}$ data, we used KaleidaGraph for optical density (OD) values between 0.02 to 0.09. Overnight-grown cultures of three biological replicates in rich LB medium or glycerol minimal medium were diluted 1:1000 and were then used for growth rate measurements. The (exponential) growth rate of the wild-type strain in a drug-free medium serves as a reference to scale all growth rates measured in this study. Raw data is reported in Tables S3 (rich LB medium) and S4 (M9-glycerol minimal medium).

**Whole genome sequencing and sequence analysis of resistant clones.** We performed whole-genome sequencing to identify the genetic basis of resistance phenotypes at different streptomycin concentrations. DNA extraction was performed on 1 mL of overnight-grown cultures by using the Epicentre DNA extraction kit and following the manufacturer's protocol. We used Illumina's Nextera XT kit to make genomic libraries (2 x 300) that were subsequently sequenced on the Illumina platform. Miseq. Samples were dual-indexed and pooled together, resulting in an average whole-genome coverage of 30X per sample. We used CLC genomics Workbench version 8 for the analysis of the fastq files obtained from Miseq sequencing to identify point mutations and small indels. To identify large deletions or transposon movements, we ran the fastq files through Breseq (version 0.27.1a) (*38*).

## 2. Fitness models for drug resistance

**Cell metabolism and growth.** Monod's law, as given in equation (1), relates the rate of exponential growth with the underlying core metabolism of the cell. To derive this relation, we assume there are two rate-limiting processes: the conversion of external nutrients into amino acids and the subsequent synthesis of proteins by ribosomes. These processes define a distribution of cellular resources: a fraction $\varphi_n$ of the proteome is allocated to amino acid synthesis and a fraction $\varphi_t = 1 - \varphi_n$ to protein synthesis. Given process rates $\kappa_n$ and $\kappa_t$, the resulting growth rate in the exponential regime takes the form

$$\lambda(\kappa_n, \kappa_t) = \max[(\Delta\varphi_n)\,\kappa_n, (\Delta\varphi_t)\,\kappa_t] \tag{5}$$

with $\Delta\varphi_n = \varphi_n - \varphi_{n,0}$ and $\Delta\varphi_t = \varphi_t - \varphi_{t,0}$, where $\varphi_{n,0}$ and $\varphi_{t,0}$ are baseline values of proteome fractions unrelated to growth. Inserting the constraint $\varphi_n + \varphi_t = 1$ shows that the maximum is attained if the flux balance condition

$$(\Delta\varphi_n)\,\kappa_n = (\Delta\varphi_t)\,\kappa_t \tag{6}$$

is fulfilled, in accordance with experimental observations (*17, 39, 40*). This condition requires a regulated ribosome content,

$$\Delta\varphi_t = C\frac{\kappa_n}{\kappa_n + \kappa_t}, \tag{7}$$

where $C = 1 - \varphi_{n,0} - \varphi_{t,0} = \Delta\varphi_{t,\max} = \lambda_{\max}/\kappa_t$ is the maximum proteome fraction of ribosomes contributing to growth. Inserting equation (7) into equation (5), we obtain Monod's law, equation (1). The Monod growth rate $\lambda(\kappa_n, \kappa_t)$ is plotted in Fig. 1b.

**Metabolism of ribosome-targeting drugs.** Here we summarize the metabolic model for ribosome-targeting drugs developed by Greulich et al. (*19, 41*) which is the basis of our metabolic fitness models. The action of the drug takes place in three steps (Fig. 1a). Given an extracellular drug level $d$, membrane transport of unbound drug molecules (with rates $\gamma_{\text{in}}$ and $\gamma_{\text{out}}$) and dilution by growth (with rate $\lambda$). Intra-cellular drug molecules bind to ribosomes (with an equilibrium constant $K$). These dynamics generate a steady-state intra-cellular drug concentration $d_{\text{int}}$, impairing a fraction $1 - p_u = 1/(1 + K/d_{\text{int}})$ of the ribosomes. The action of the drug results in a growth reduction

$$\frac{\lambda}{\lambda_0} = g\left(\frac{d}{d_*}, \frac{\lambda_0}{\lambda_*}\right), \tag{8}$$

depending on the drug response parameters

$$\lambda_* = 2\sqrt{\gamma_{\text{out}}\,\kappa_t^0\,K}, \qquad d_* = \frac{C\,\lambda_*}{2\gamma_{\text{in}}}, \tag{9}$$

where the metabolic constants $\kappa_t^0$ and $C = \lambda_{\max}/\kappa_t^0$ refer to drug-free growth. Introducing the scaled variables $\delta = d/d_*$ and $r = \lambda_0/\lambda_*$, the drug response curve $g(\delta, r)$ can most easily be written in terms its inverse function,

$$\delta(g, r) = \frac{1}{2rg}(1 - g + 4\,r^2 g^2 - 4\,r^2 g^3), \tag{10}$$

which specifies the scaled drug level $\delta$ required for a growth reduction $g$. In particular, the half-inhibitory concentration is given by

$$\frac{d_{50}}{d_*} = \frac{1}{2}\left(r + \frac{1}{r}\right). \quad (11)$$

This model shows that drug action is a perturbation of the cell's core metabolism that is shaped by two important nonlinearities. First, the dilution of intra-cellular drug levels by cell divisions introduces a negative feedback between drug level and growth, which can generate fitness cliffs with a sudden drop in growth at a threshold drug level (*18, 19, 42*). Second, under a drug-induced reduction of the translational capacity, $\kappa_t(d) = p_u \kappa_t^0$, cells upregulate their ribosome content according to equation (7), mitigating the drug effect on growth. From equations (1) and (7), the drug-dependent proteome fraction of ribosomes can be written in the form

$$\Delta\varphi_t(d, q) = C\frac{q + 1 - g}{q + 1}, \quad (12)$$

where $q = \kappa_n/\kappa_t$ is the ratio of metabolic rates in a drug-free environment and $g$ is given by equation (10). To display the contribution of ribosome regulation to drug defense, we evaluate growth curves of a hypothetical cell with its ribosome content constrained to the drug-free value, $\Delta\varphi_t = Cq/(1 + q)$. Constrained cells violate the flux balance condition (6) in the presence of the drug. To introduce the constraint into the metabolic model of drug action, we self-consistently determine a metabolic ratio $\tilde{q}$, such that $\Delta\varphi_t(d, \tilde{q}) = Cq/(1 + q)$. A straightforward calculation then determines the growth curve $g_{\text{ur}}(\delta, q, r)$ of the constrained cell, which can again be written in terms its inverse function,

$$\delta_{\text{ur}}(g, q, r) = \delta\left(\frac{gq + 1}{q + 1}, \frac{g(q + 1)}{gq + 1}r\right) \quad (13)$$

with $\delta(.,.)$ given by equation (10). Fig. 4a shows the growth curve $g_{\text{ur}}(d/d_*, q^{\text{wt}}, r^{\text{wt}})$ for our wild type. Ribosome regulation is seen to partially compensate the drug effect on growth in the sub-lethal dosage regime. In particular, $d_{50}^{\text{wt}}$ is significantly higher than for the constrained cell.

**Fitness model for drug resistance evolution.** We assume that a given mechanism of drug resistance generates a quantitative molecular resistance trait and a family of mutants with different trait amplitude $\varepsilon$. We also assume that the resistance mutants follow the same metabolic model of drug action as the wild type and have computable changes in core metabolism (with parameters $\kappa_n$, $\kappa_t$, and $C$) and drug interactions (with parameters $\gamma_{\text{in}}$, $\gamma_{\text{out}}$, and $K$). This input determines a fitness model specifying the growth of mutants with a given resistance amplitude at a given drug level, $\lambda/\lambda_0^{\text{wt}} = G(\varepsilon, d/d_*^{\text{wt}}; q^{\text{wt}}, r^{\text{wt}})$. The fitness model takes the general form

$$G(\varepsilon, d) = W(\varepsilon)\, g\left(\frac{1}{D(\varepsilon)}\frac{d}{d_*^{\text{wt}}}, \frac{W(\varepsilon)}{L(\varepsilon)}r^{\text{wt}}\right) \quad (14)$$

with transformation functions

$$W(\varepsilon) = \frac{\lambda_0(\varepsilon)}{\lambda_0^{\text{wt}}}, \qquad D(\varepsilon) = \frac{d_*(\varepsilon)}{d_*^{\text{wt}}}, \qquad L(\varepsilon) = \frac{\lambda_*(\varepsilon)}{\lambda_*^{\text{wt}}} \quad (15)$$

quantifying the effects of the resistance trait on drug-free growth and on drug response. Here we have temporarily suppressed the dependence of the functions $G, W, D, L$ on the wild type parameters $q^{\text{wt}}, r^{\text{wt}}$. For a given resistance mechanism, the fitness model of equations (14) and

(15) establishes a computable relation between evolutionary changes, drug and nutrient conditions, and growth. Specific forms of this model for the different resistance mechanisms discussed in this paper will be given below.

**Resistance trade-off function.** For a given resistance mechanism, the fitness model of equations (13) and (14) predicts a unique relation between the level of resistance, $R = d_{50}/d_{50}^{\text{wt}}$, and the corresponding cost in a drug-free medium, $W = \lambda_0/\lambda_0^{\text{wt}}$. This tradeoff can be computed in parametric form, using the relations

$$W(\varepsilon), \qquad R(\varepsilon) = \frac{r^{\text{wt}} D(\varepsilon)}{(1+(r^{\text{wt}})^2)} \left( \frac{W(\varepsilon)}{L(\varepsilon)} r^{\text{wt}} + \frac{L(\varepsilon)}{W(\varepsilon)} \frac{1}{r^{\text{wt}}} \right), \qquad (16)$$

where $R(\varepsilon)$ has been obtained from equations (11) and (15). Substituting the inverse function $\varepsilon(R)$ into $W(\varepsilon)$ then gives the explicit relation $W(R)$. Tradeoff functions of this form are shown in Fig. 2 and Fig. S4.

**Maximum-growth and critical resistance trajectories.** For a given resistance mechanism, the fitness model also predicts the maximum growth rate of mutants as a function of the drug level,

$$G_c(d) = \max_\varepsilon G(d, \varepsilon). \qquad (17)$$

Hence the maximum-growth rate is obtained as the envelope of the family of growth curves of mutants with a common resistance mechanism, parameterized by the trait amplitude $\varepsilon$. To evaluate the condition (17), we use again the inverse function $d/d_{50}^{\text{wt}} = \Delta(G, \varepsilon)$, where

$$\Delta(G, \varepsilon) = D(\varepsilon) \delta \left( \frac{G}{W(\varepsilon)}, \frac{rW(\varepsilon)}{L(\varepsilon)} \right)$$

is given by equations (10), (14), and (15). This determines the maximum growth trajectory in parametric form. The growth rate $G_c(\varepsilon)$ is obtained from the condition $\partial_\varepsilon \Delta(G, \varepsilon) = 0$, which amounts to finding the largest positive solution of the cubic equation

$$(2r^{\text{wt}})^2 LWD\partial_\varepsilon \log\left(\frac{LW}{D}\right) G_c^3 + (2r^{\text{wt}})^2 LW^2 D\partial_\varepsilon \log\left(\frac{D}{L}\right) G_c^2 - L^3 WD\partial_\varepsilon \log\left(\frac{DL}{W}\right) G_c + L^3 W^2 D\partial_\varepsilon \log(DL) = 0. \qquad (18)$$

The corresponding drug level is given by

$$\frac{d_c(\varepsilon)}{d_{50}^{\text{wt}}} = \Delta(G_c(\varepsilon), \varepsilon) = \frac{D(\varepsilon)(W(\varepsilon) - G_c(\varepsilon))\left(L^2(\varepsilon) - (2r^{\text{wt}} G_c(\varepsilon))^2\right)}{(1+(r^{\text{wt}})^2) G_c(\varepsilon) L(\varepsilon) W(\varepsilon)}. \qquad (19)$$

Maximum-growth trajectories are shown in Fig. 3a, 4a, and S4. In a similar way, the critical resistance trajectory $R_c(d)$, which gives the resistance level of maximum growth-mutants as a function of the drug-level, can be evaluated in parametric form, using the relations (16) and (19). Critical resistance trajectories are shown in Fig. 3bc. We emphasize that these maximum growth trajectories emerge for any evolutionary mechanism with a continuous effect parameter acting upon a metabolic growth law, as given by equations (10), (14), and (15).

**Variation of nutrient conditions.** Fitness depends not only on drug level but also on nutrient conditions. Specifically, the nutrient level sets the wild type nutritional capacity relative to a reference medium,

$$\nu = \frac{\kappa_n^{\text{wt}}}{\kappa_n^{\text{ref}}} = \frac{q^{\text{wt}}}{q^{\text{ref}}}, \tag{20}$$

which, in turn, determines the relation of drug-free growth rates,

$$\frac{\lambda_0^{\text{wt}}}{\lambda_0^{\text{ref}}} = \frac{\nu(q^{\text{ref}} + 1)}{\nu q^{\text{ref}} + 1}. \tag{21}$$

The fitness model (14) then predicts how mutant growth rates jointly depend on drug and nutrient level,

$$\frac{\lambda}{\lambda_0^{\text{ref}}} = \frac{\nu(q^{\text{ref}} + 1)}{\nu q^{\text{ref}} + 1} G\left(\varepsilon, d; \nu q^{\text{ref}}, \frac{\nu(q^{\text{ref}} + 1)}{\nu q^{\text{ref}} + 1} r^{\text{ref}}\right). \tag{22}$$

This transformation uses the assumption that the drug response rate $\lambda_*$ is an intrinsic parameter independent of the environment. In Fig. 4d, we use equation (22) to predict growth curves of membrane permeability mutants in glycerol minimum medium, using rich medium as a reference (see also Data Analysis below).

**Specific resistance mechanisms.** In the following, we specify the fitness model of equations (14), (15) for the quantitative resistance traits discussed in the main text.

(a) **Membrane permeability.** The minimal model of membrane-based resistance evolution is defined by a coupled reduction of drug and nutrient uptake with a permeability parameter $\varepsilon < 1$, as given by equation (2) of the main text. Using equation (9), we obtain a fitness model with transformation functions

$$D(\varepsilon) = \frac{1}{\varepsilon}, \qquad L(\varepsilon) = 1, \qquad W(\varepsilon) = \frac{\varepsilon(q^{\text{wt}} + 1)}{\varepsilon q^{\text{wt}} + 1}. \tag{23}$$

The resulting maximum-growth trajectories, as given by equations (20) and (21), take the simple form

$$G_c(\varepsilon) = \frac{\varepsilon(q^{\text{wt}} + 1)}{\varepsilon q^{\text{wt}} + 2}, \qquad \frac{d_c(\varepsilon)}{d_{50}^{\text{wt}}} = \frac{\left(q^{\text{wt}}(1 - G_c(\varepsilon)) + 1\right)^2 ((2r^{\text{wt}})^2 G_c^2(\varepsilon) + 1)}{4(1 + (r^{\text{wt}})^2)(q^{\text{wt}} + 1) G_c^2(\varepsilon)}, \tag{24}$$

which is used in the main text. Results from this model are plotted in Fig. 2, 3, and 4. The growth curves in minimal glycerol are obtained by using the transformation (22) with parameter $\nu = q^{\text{wt}}/q^{\text{ref}} = 0.18 / 5.9$ (Fig. 4d). We note that the membrane permeability model applies to any molecular uptake mechanism that affects drug and nutrients proportionally, as given by equation (2). In contrast, the *Cpr* stress response system, which is observed in minimal glycerol, affects drug and nutrients via different pathways. Hence, the proportionality constraint, equation (2), no longer applies, which is consistent with an observed growth curve $G(d)$ outcompeting the prediction of equation (24).

We also consider an extended membrane model with proportional reduction of the drug uptake and release rates,

$$\varepsilon = \frac{\gamma_{\text{in}}}{\gamma_{\text{in}}^{\text{wt}}} = \frac{\gamma_{\text{out}}}{\gamma_{\text{out}}^{\text{wt}}} = \frac{\kappa_n}{\kappa_n^{\text{wt}}} < 1, \tag{25}$$

which has the transformation functions

$$D(\varepsilon) = \frac{1}{\sqrt{\varepsilon}}, \qquad L(\varepsilon) = \sqrt{\varepsilon}, \qquad W(\varepsilon) = \frac{\varepsilon(q^{\text{wt}} + 1)}{\varepsilon q^{\text{wt}} + 1}. \tag{26}$$

Under irreversible drug metabolism, which applies our regime of membrane-based resistance evolution (Fig. 4), the extended membrane model produces similar growth curves as the minimal model (Fig. S4a), as expected from the drug action model of ref. (*19*). For comparison, Fig. S4b shows growth curves of the minimal membrane model under reversible drug metabolism, which applies, for example, to the aminoglycosides tetracycline (*43*) and chloramphenicol (*44*).

(b) **Efflux pumps.** Our minimal model describes an increase in the drug efflux rate $\gamma_{\text{out}}$ by a constitutive over-expression of efflux pumps at a proteome fraction $\varphi_{\text{efl}}$,

$$\varepsilon = \frac{\gamma_{\text{out}}}{\gamma_{\text{out}}^{\text{wt}}} = \frac{\varphi_{\text{efl}}}{\varphi_{\text{efl}}^{\text{wt}}} > 1. \tag{27}$$

This over-expression reduces the maximum growth rate (*17, 45*) and causes a fitness cost $C/C^{\text{wt}} = \Delta\varphi_{t,\max}/\Delta\varphi_{t,\max}^{\text{wt}} = 1 - (\varepsilon - 1)c_{\text{efl}}$; the cost parameter $c_{\text{efl}} = (1/C)(\partial C/\partial \varepsilon) = \varphi_{\text{efl}}^{\text{wt}}/\Delta\varphi_{t,\max}$ measures the fraction of the growth-related proteome invested for the expression of efflux pumps and other constitutive cost factors, such as energy consumption.

Together with equation (9), we obtain the transformation functions

$$D_{\text{efl}}(\varepsilon) = \sqrt{\varepsilon}\, W_{\text{efl}}(\varepsilon), \qquad L_{\text{efl}}(\varepsilon) = \sqrt{\varepsilon}, \qquad W_{\text{efl}}(\varepsilon) = 1 - (\varepsilon - 1)c_{\text{efl}}. \tag{28}$$

The resulting maximum-growth trajectories $G_{c,\text{efl}}(d)$ are given in parametric form by the relations

$$G_{c,\text{efl}}(\varepsilon), \qquad \frac{d_{c,\text{efl}}(\varepsilon)}{d_{50}^{\text{wt}}} = \frac{1 + c_{\text{efl}} - G_c + 4c_{\text{efl}}(r^{\text{wt}} G_c)^2}{4((r^{\text{wt}})^2 + 1)c_{\text{efl}} G_c}, \tag{29}$$

where $G_{c,\text{efl}}$ is the positive solution of

$$(2r^{\text{wt}})^2 c_{\text{efl}} G_{c,\text{efl}}^2 + G_{c,\text{efl}} + c_{\text{efl}}(2\varepsilon - 1) - 1 = 0.$$

In the intermediate drug regime ($d/d_{50}^{\text{wt}} < 5$) and under irreversible drug metabolism ($r^{\text{wt}} \gg 1$), these trajectories remain below those of membrane mutations. Hence, in this regime, efflux pumps are predicted to be a less efficient resistance mechanism than membrane permeability reduction in tune with the absence of efflux mutants in our Luria-Delbrück experiments. Here we use an order-of-magnitude estimate of the cost parameter, $c_{\text{efl}} \sim 10^{-2}$, which can be inferred from measurements of the fitness cost of efflux pumps (*3, 46, 47*). The minimal model makes a conservative assumption: the fitness cost of efflux pumps arises primarily by changes in proteome resource allocation due to their constitutive over-expression. Other cost factors, including load-dependent energy consumption of pumps and loss of nutrients by efflux, can shift the fitness balance of efflux pumps.

**Evolutionary switches between resistance mechanisms.** As discussed above, the wild type partially offsets the action of ribosome-targeting drugs by upregulation of its ribosome content according to equation (12). Given the evolutionary cost $W$ of resistance mutations, the regulated wild-type is always the maximum-fitness strain at sufficiently low drug levels. Our fitness model predicts a switch from regulation to membrane-based resistance evolution. The onset of evolution occurs at a drug level $d_{\text{rm}}(q^{\text{wt}}, r^{\text{wt}})$ determined by the condition

$$g(d_{\text{rm}}, r^{\text{wt}}) = G_c(d_{\text{rm}}, q^{\text{wt}}, r^{\text{wt}}), \qquad (30)$$

which compares the wild-type growth curve and the maximum-growth trajectory of membrane evolution, equations (8) and (17). The resulting crossover point is given in parametric form,

$$G_{\text{rm}}(q^{\text{wt}}) = G_c(\varepsilon = 1, q^{\text{wt}}) = \frac{q^{\text{wt}} + 1}{q^{\text{wt}} + 2}, \qquad d_{\text{rm}}(q^{\text{wt}}, r^{\text{wt}}) = d_c(\varepsilon = 1, q^{\text{wt}}, r^{\text{wt}}), \quad (31)$$

which can be evaluated from equation (18) (Fig. 4a). The predicted crossover for streptomycin in rich LB is at $d_{\text{rm}} = 0.43 \, d_{50}^{\text{wt}}$, in qualitative agreement with previously observed onset of adaptive resistance evolution at sub-MIC levels (*9*). More generally, our model can be used to compare arbitrary resistance mechanisms. Given two mechanisms A, B with maximum-growth curves $G_{c,A}$ and $G_{c,B}$, a selective switch between these mechanisms is predicted at a crossover point $d_{AB}$ given by $G_{c,A}(d_{AB}, q^{\text{wt}}, r^{\text{wt}}) = G_{c,B}(d_{AB}, q^{\text{wt}}, r^{\text{wt}})$.

## 3. Data analysis

**Inference of model-based growth curves.** For the wild type and each membrane mutant, we independently fit the experimental growth data to growth curves of the form

$$\lambda(d) = \lambda_0 g\left(\frac{d}{d_*}, \frac{\lambda_0}{\lambda_*}\right), \qquad (32)$$

where $\lambda_0$ is the drug-free growth rate and $g(\delta, r)$ is a drug response curve of the form (10), as given by the metabolic model of Greulich et al. (*19*). The fitted growth curves are plotted in Fig. S2 together with the growth measurements. Each fit involves three independent parameters: the rate $\lambda_0$ and the drug response parameters $d_*, \lambda_*$. We note that in the regime of low expected growth ($\lambda/\lambda_0^{\text{wt}} \lesssim 0.1$), the correspondence between data and model is confounded by two effects: mutants in this regime are unlikely to reach the detection threshold in Luria-Delbrück assays, and new metabolic effects an invalidate quantitative model predictions (*49*). However, the fit parameters are determined predominantly by data points outside this regime, leading to a robust inference procedure.

Model growth curves are gauged by an error score $S = \sum_i E(\lambda_i, \hat{\lambda}_i)$, where the index $i$ runs over all measurements of a given growth curve, $\lambda_i$ is the measured growth rate, and $\hat{\lambda}_i$ is the corresponding model rate. We use an error function $E(\lambda, \hat{\lambda}) = (\lambda - \hat{\lambda})^2/\sigma^2$ for $\lambda > \lambda_{\min}$ and $E(\lambda, \hat{\lambda}) = (\max(\lambda, \lambda_{\min}) - \lambda_{\min})^2/\sigma^2$ for $\lambda \leq \lambda_{\min}$, taking into account measurement errors of order $\lambda_{\min} = 0.03 \, \lambda_0^{\text{wt}}$ preventing detection of small growth rates. We use an expected square error of the form $\sigma^2 = \max(\text{Var}(\lambda), \lambda_{\min}^2)$, where $\text{Var}(\lambda)$ is the recorded variance of replicate measurements. Numerical minimization uses simplex search by the Nelder-Mead algorithm (*48*) (we check convergence with varying initial conditions). To obtain these distributions we perform 1000 fits to synthetic growth inhibition curves for each mutant and extract the corresponding optimal parameters. Synthetic growth curves are sampled from distributions defined by the experimental data points: Above the threshold $G_{\min}$, we use Gaussian distributions with mean and standard deviation given by the measurements in biological replicates. If no growth is reported, we sample from a uniform distribution in the interval $[0, G_{\min}]$. This posterior distribution determines the posterior average values, which turn out to be very close to the corresponding maximum-likelihood values, and the 90% confidence intervals reported in Table S2. Raw experimental data is given in Tables S3 and S4.

In rich medium, we infer the wild type parameters $\lambda_0^{\text{wt}} = 2.02$ (1.98, 2.10)/hr, $d_*^{\text{wt}} = 3.13$ (2.98, 3.23) mg/L, and $\lambda_*^{\text{wt}} = 0.37$ (0.35, 0.40)/hr (posterior average value, 90% confidence interval), resulting in the dimensionless drug response parameter $r^{\text{wt}} = 5.42$ (5.21, 5.60). The inferred parameters $\lambda_0, d_*, \lambda_*$ for all membrane mutants are reported in Table S2 in units of the corresponding wild type parameters; we define the scaled drug-free growth rate $W = \lambda_0/\lambda_0^{\text{wt}}$. The fitted growth curves also produce inferred values of the membrane permeability, $\varepsilon = \gamma_{\text{in}}/\gamma_{\text{in}}^{\text{wt}} = \lambda_* d_*^{\text{wt}}/(\lambda_*^{\text{wt}} d_*)$, and the drug release rate, $\gamma_{\text{out}}/\gamma_{\text{out}}^{\text{wt}} = (\lambda_*/\lambda_*^{\text{wt}})^2$, for all mutants; see equation (9) (Table S2).

The wild type parameter $q^{\text{wt}} = \kappa_n^{\text{wt}}/\kappa_t^{\text{wt}}$ in rich medium is obtained independently from the spectrum of drug-free mutant growth rates, $W(\varepsilon)$, as given by equation (3). The empirical spectrum is of the Michaelis-Menten form, equation (1), if the permeability parameter is re-interpreted as the nutritional capacity relative to the wild type, $\varepsilon = \kappa_n/\kappa_n^{\text{wt}}$. This identification is consistent under the assumption that membrane mutations have no effect on translation, i.e., $\kappa_t = \kappa_t^{\text{wt}}$ for all mutants. The Michaelis-Menten fits produce the inferred parameter $q^{\text{wt}} = 5.9$ (5.5, 6.3); the maximum-likelihood fit is shown in Fig. 2. While $\kappa_n^{\text{wt}}$ and $\kappa_t^{\text{wt}}$ can be inferred from solely from wild-type (RNA and proteomics) data (*17*), the evolutionary cost pattern $W(\varepsilon)$ provides an alternative, simple way of extracting the ratio $q^{\text{wt}}$ that does not require expression data.

The experiments in minimal glycerol medium operate at significantly reduced growth rates ($\lambda \leq \lambda_0^{\text{wt}} = 0.39$/hr). In the low-growth regime ($\lambda \lesssim 0.2$/hr), we observe a different growth pattern with an extended lag phase, and metabolic differences in ribosome function and utilization have been reported at low growth rates (*49*). Here we obtain a model-based growth curve for the wild-type by equation (22), constraining $\lambda_*^{\text{wt}}$ to its value in rich medium and allowing an independent fit of $d_*^{\text{wt}}$. This parameter can change, in particular, with metabolic shifts affecting the growth-related proteome fraction, $C$; see equation (7). Fitting this model to the points of observed positive growth, we obtain $d_*^{\text{wt}} = 2.0$ mg/L and $\lambda_0^{\text{wt}} = 0.39$/hr, corresponding to $q^{\text{wt}} = 0.2$ and $r^{\text{wt}} = 1.05$; the resulting dosage response curve is shown in Fig. 4b and Fig. S2b. In the low-growth regime, model-based growth curves should be regarded as upper bounds to observed growth rates.

**Fit quality and comparison of membrane models.** Fitting the metabolic drug response model (*19*) to our wild type and membrane mutant growth curves produces a total error score $S = 130$ for a total of 150 independent measurements in 24 strains. The score per measurement is of order 1, which implies the metabolic model reproduces the empirical data within measurement errors (Fig. S2). We can also compare the full membrane model with an alternative model with the global constraint $\lambda_* = \lambda_*^{\text{wt}}$, which amounts to fixing the drug release rate, $\gamma_{\text{out}}$, of all mutants to the wild type value. The constrained model produces a fit of comparable quality, which has a total error score $S = 154$ for a total of 130 independent measurements in 21 strains (excluding the outliers in Fig. S3). We conclude that variation of $\gamma_{\text{out}}$ across mutants is negligible for most growth curves of our membrane mutants (see also Fig. S3). This asymptotic independence is expected in the so-called irreversible drug response regime ($r \gg 1$) described in ref. (*19*).

**Validation of the fitness model.** The empirical growth curve of a given membrane mutant has two characteristic drug levels that are important for testing the membrane fitness model.

(i) The half-inhibitory concentration, $d_{50}$, has the simple form $d_{50} = (d_*/2)(\lambda_0/\lambda_* + \lambda_*/\lambda_0)$ in terms of the basic growth parameters $\lambda_0, d_*, \lambda_*$ obtained from our fitting procedure; see equation (11). The resulting resistance values $R = d_{50}/d_{50}^{\text{wt}}$ are reported in Table S2 for all membrane mutants. The empirical parameter $R$ depends solely on the drug response curve $g$ of a given mutant; a priori, the drug-free growth $W = \lambda_0/\lambda_0^{\text{wt}}$ can vary independently across the set of membrane mutants. Our fitness model establishes a unique relation $W(R)$, which is given in parametric form by equation (3). This relation is found to be in quantitative agreement with the empirical pattern (Fig. 2c).

(ii) The critical drug concentration, $d_c$, is the point where a given mutant is predicted to confer maximum fitness against all other membrane mutants. This point is defined by equation (24) solely in terms of the mutant permeability parameter, $\varepsilon = \gamma_{\text{in}}/\gamma_{\text{in}}^{\text{wt}}$, and the wild type parameters $q^{\text{wt}}, r^{\text{wt}}$. Our fitness model predicts the growth rate of that mutant at the critical drug concentration, $G_c(\varepsilon)$, by equation (4). This relation is validated in Fig. 3a. Notably, the critical growth rate becomes independent of the parameter $r^{\text{wt}}$.

**Inference of mutation rates.** Our Luria-Delbrück assays in rich agar record a drug-dependent establishment rate of membrane mutants, $u_m(d)$, which strongly decreases with increasing $d$ in the intermediate-dosage regime ($d/d_{50}^{\text{wt}} > 2$) (Fig. 4a, Table S5). We estimate an effect-dependent spectrum of membrane mutations, using the heuristic form $U_m(R) \sim u_0/(1 + \exp[a(R - R_0)])$ and the relation $U_m(d) = \int U(R)\,\theta(G(d, \varepsilon(R)) - G_{\text{thr}})\,dR$, where $G_{\text{min}} \approx 0.1$ is the minimum relative growth rate giving rise to detectable colonies after an overnight culture and $G(d, \varepsilon)$ is the growth rate given by our fitness model. The fitted mutational spectrum has parameters $u_0 = 3.6 \times 10^{-7}$, $a = 3.4, R_0 = 2.7$ (Fig. S5a).

Target mutations appear in rich agar in the dosage regime $d/d_{50}^{\text{wt}} \geq 5.5$ with establishment rates $U_t$ of order $10^{-10}$ (Fig. 4a). We observe ten genetic variants, which have the amino acid substitutions P91L, P91Q, P91R, G92D, K43T, K43N, K43R, K43Q, K88R, K88E in the *rpsL* gene. Growth measurements of three prevalent variants, P91L, K43N, K43T, and the fittest variant K43R, reveal approximately drug-independent growth rates $G = 0.36, 0.76, 0.85, 0.95$ for $d/d_{50}^{\text{wt}} \geq 5.5$. Establishment rates for individual variants are estimated from observed frequencies of mutant colonies (Fig. S5a). In minimal glycerol, target mutations have establishment rates of order $10^{-10}$ (Fig. 4c). We observe 4 genetic variants with substitutions K88R, K43T, K43N, and K43R in the *rpsL* gene, all of which have approximately drug-independent growth rates $G > 0.9$ for $d/d_{50}^{\text{wt}} \geq 1.9$. Stress response mutations in the gene *cpxA* are observed only in minimal glycerol at $d/d_{50}^{\text{wt}} = 0.9$ with an establishment rate $U_{cpx} \sim 10^{-8}$ (Fig. 4c).

These mutational spectra are used as input to determine the growth rate spectrum at a given drug level, $U(G; d)$, which governs short-term evolution experiments (Fig. S5b), as well as the normalized distributions $P_{\text{LD}}(R; d)$ and $P_{\text{LD}}(G; d)$ relevant for Luria-Delbrück assays (Fig. S5cd).

**Additional references**

# Supplementary Figures and Tables

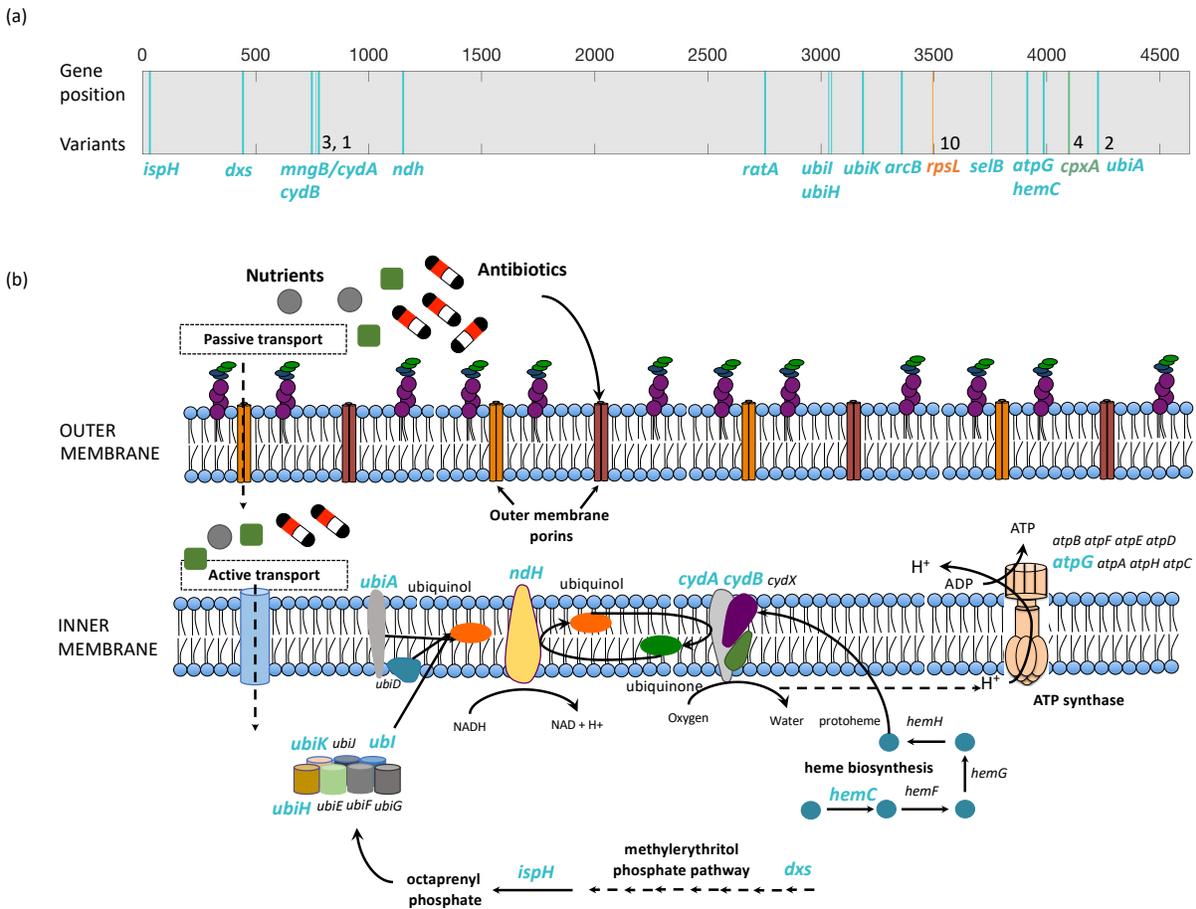

**Fig. S1: Genomics and mechanisms of resistance evolution.** (a) Genomic distribution of resistance mutations found in the Luria-Delbrück experiments of this study. Bars indicate the genomic coordinates of affected genes, multiple distinct mutations in a given gene are marked by numbers. Functional classes of these mutations include membrane active transport (cyan, 17 distinct genome mutations in 23 clones grown in rich medium), drug target modifications (orange, 10 distinct amino acid changes in the *rpsL* gene in rich medium, 4 in minimal glycerol medium, and *Cpx* stress response (green, 4 distinct amino acid changes in the gene *cpxA* in minimal glycerol medium). These functional classes account for 94% of the resistance mutants found in this study; all mutations are listed in Tables S3 and S4. (b) Schematic representation of key pathways involved in active membrane transport (mutated genes in brackets and marked in cyan): ubiquinone biosynthesis (*ubiA, ubiI, ubiH, ubiK*), ATP biosynthesis (*atpG*), heme biosynthesis (*hemC*), methylerythritol phosphate pathway (*dxs, ispH*), NADH-cytochrome oxidase electron transfer (*ndh, cydA, cydB*); see Table S1.

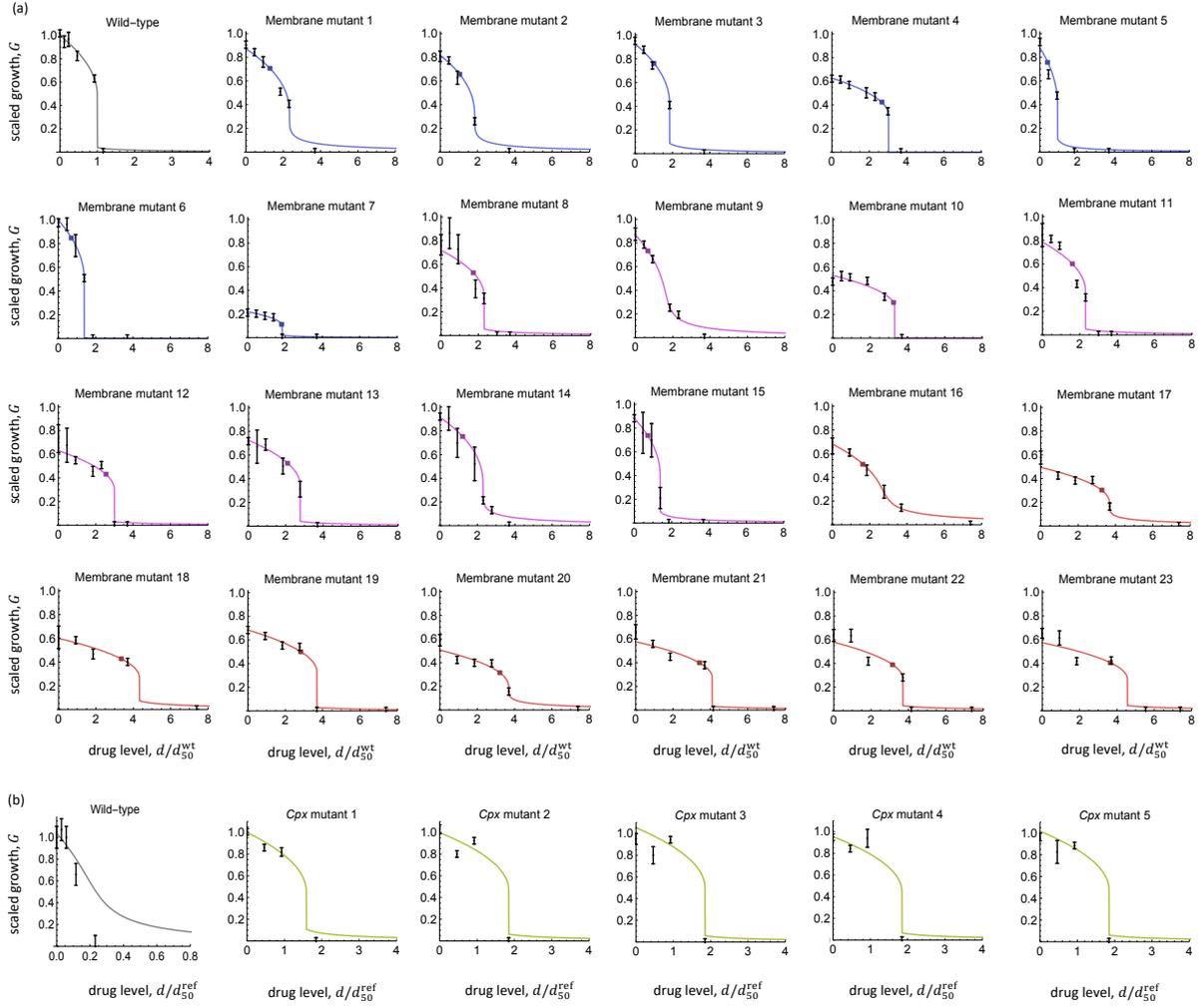

**Fig. S2: Drug-dependent growth curves.** (a) Rich liquid medium. Data points show growth rates of the wild type and of membrane mutants at different drug levels $d$ (measured in units of the half-inhibitory concentration of the wild type, $d_{50}^{\text{wt}} = 8.7\,\text{mg/L}$). Bars indicate rms. experimental uncertainties, colors mark the drug level of the Luria-Delbrück assay used to elicit each mutant, $d_{\text{LD}}/d_{50}^{\text{wt}} = 0.9, 1.9, 3.6$ (violet, pink, red). Empirical growth curves, $\lambda(d)/\lambda_0^{\text{wt}} = G(d; W, d_{50}^*, \lambda_0^*)$ (lines) involve three independent fit parameters for each mutant: the drug-free growth rate, $W = \lambda_0/\lambda_0^{\text{wt}}$, and the drug response parameters (*19*) $d_{50}^*, \lambda_0^*$; see equation (9). These fits also produce estimates of the membrane transport rates, $\gamma_{\text{in}}$ and $\gamma_{\text{out}}$, and of the characteristic drug levels $d_c$ and $d_{50}$. For each mutant, the critical point $(d_c, G(d_c))$ (square) gives the empirical growth rate at the predicted critical drug concentration; this point is used in Fig. 3a. Inferred growth and resistance parameters for all membrane mutants are listed in Table S2, raw data are reported in Table S3. (b) Minimal glycerol liquid medium. Data points show growth rates of the wild type and of *cpx* stress response mutants elicited at $d_{\text{LD}}/d_{50}^{\text{ref}} = 0.9$. All drug levels are measured in units of $d_{50}^{\text{ref}} = 8.7\,\text{mg/L}$. The fit procedure is detailed in Methods.

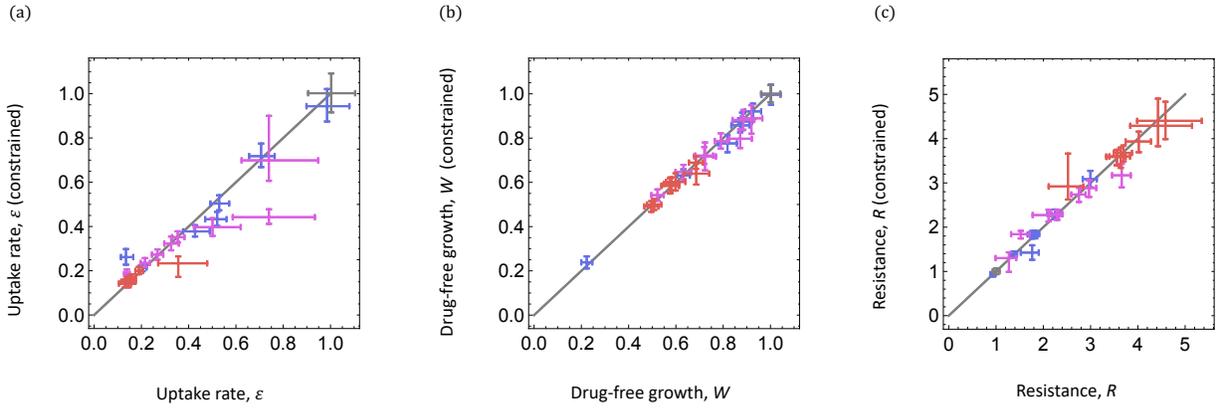

**Fig. S3: Membrane model of drug resistance.** We compare (a) uptake rate, $\varepsilon = \gamma_{in}/\gamma_{in}^{wt}$, (b) drug-free growth rate, $W = \lambda_0/\lambda_0^{wt}$, and (c) resistance, $R = d_{50}/d_{50}^{wt}$, of membrane mutants obtained from our full inference procedure with the corresponding values from a constrained model with fixed parameter $\lambda_* = 2(\gamma_{out} \kappa_t^0 K)^{1/2} = \lambda_*^{wt}$. Drug-free growth and resistance show insignificant changes, the uptake rate changes significantly in only three mutants. Hence, variation of the parameter $\gamma_{out}$ does not affect the inference of the membrane model (equation (2)), of the evolutionary tradeoff $W(R)$ (Fig. 2), and of the empirical data reported in Fig. 3.

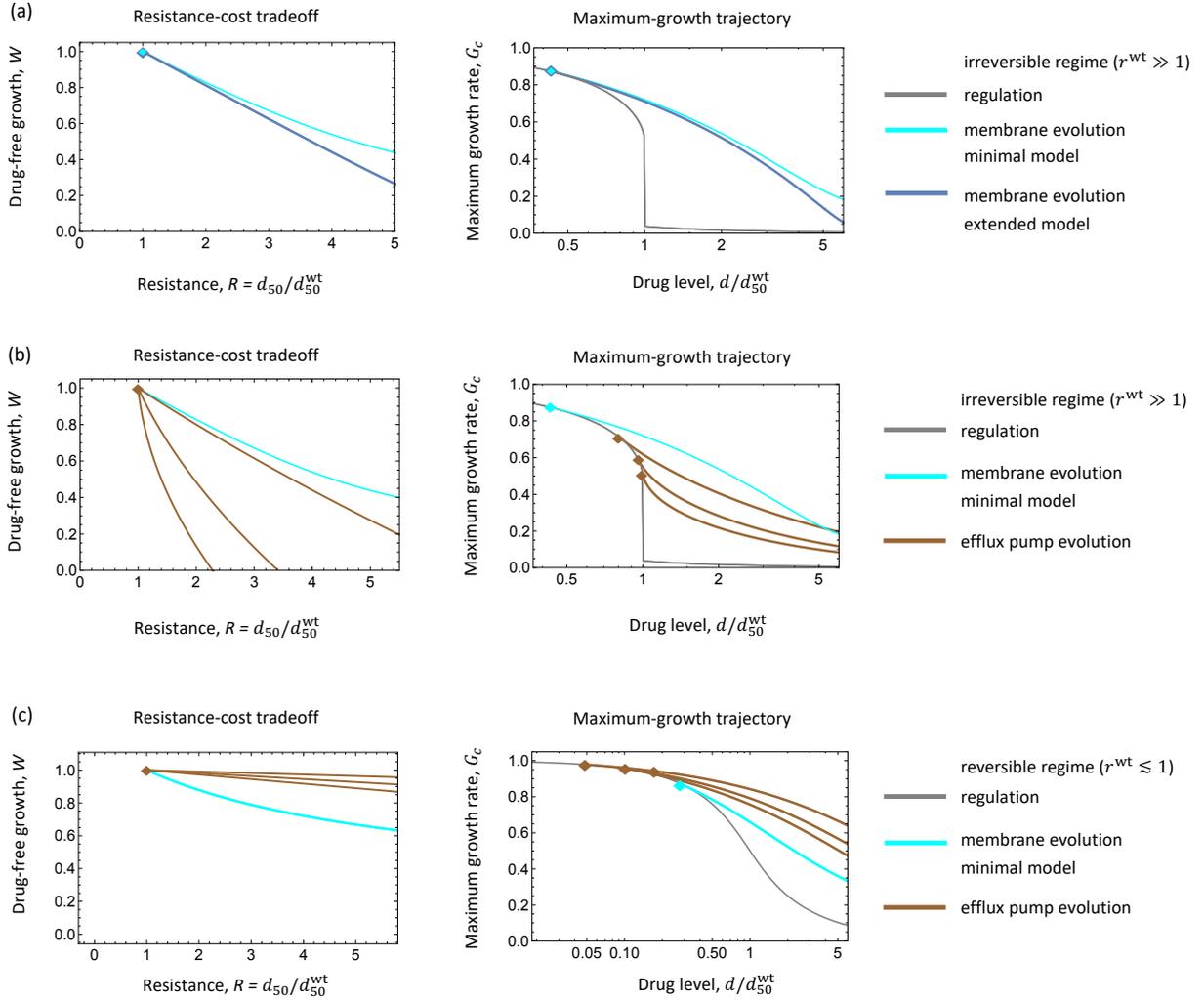

**Fig. S4: Comparison of evolutionary resistance mechanisms.** Evolutionary tradeoff curves, $W(R)$, and maximum-growth trajectories, $G_c(d)$, are shown for the following models: **(a)** Minimal membrane permeability evolution (reduction of uptake rates, $\gamma_{in}/\gamma_{in}^{wt} = \kappa_n/\kappa_n^{wt}$, at constant release rate, $\gamma_{out}/\gamma_{out}^{wt} = 1$, as in main text) and an extended model (reduction of uptake and release rates, $\gamma_{in}/\gamma_{in}^{wt} = \kappa_n/\kappa_n^{wt} = \gamma_{out}/\gamma_{out}^{wt}$) are compared in the growth regime of irreversible drug metabolism (19) ($r^{wt} \gg 1$). In this regime, release rates have a negligible influence on growth and resistance, supporting use of the minimal model. Model parameters: $q^{wt} = 5.9$, $r^{wt} = 5.4$ as in main text. **(b, c)** Evolution of drug efflux pumps (increase of drug release rate by overexpression of efflux genes, $\gamma_{out}/\gamma_{out}^{wt} = \varphi_{efl}/\varphi_{efl}^{wt}$) and minimal membrane evolution are compared in regimes of irreversible ($r^{wt} \gg 1$) and reversible growth ($r^{wt} \lesssim 1$). Efflux pumps are predicted to be relatively inefficient specifically under irreversible growth. Model parameters: efflux cost parameter, $c_{efl} = 5 \times 10^{-3}, 1 \times 10^{-2}, 1.5 \times 10^{-2}$ (top to bottom), $q^{ref} = 5.9$, $r^{ref} = 5.4$ (irreversible regime, as in main text), $q^{wt} = 5.9$, $r^{wt} = 0.9$ (reversible regime). The reversible regime can be attained by applying a drug with reduced ribosome binding affinity (i.e., with increased equilibrium constant $K$) for a given wild-type (i.e., at constant $\lambda_0^{wt}$). This results in increased drug response parameters $\lambda_*^{wt}$ and $d_*^{wt}$ compared to the reference drug; see equation (9).

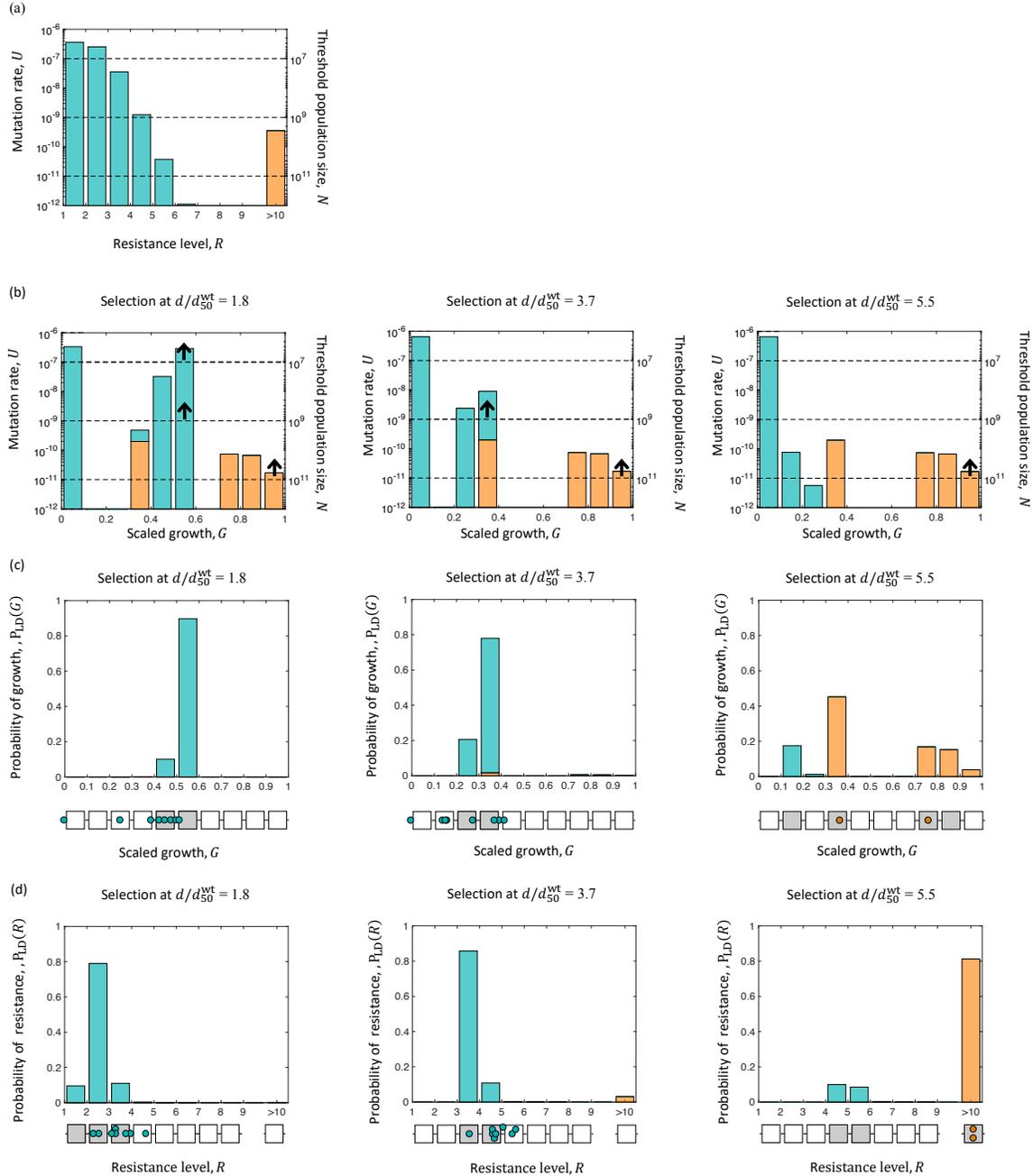

**Fig. S5: Resistance mutation spectra in evolution and selection assays.** (a) Spectrum of resistance mutation rates, $U(R)$, inferred from Luria-Delbrück assays (cyan: membrane mutations; orange: *rpsL* mutations). Dashed horizontal lines indicate threshold population sizes; resistance effects above a given line are likely to be represented in a population of given size $N$. (b) Resulting spectrum of mutant growth rates in rich LB at different drug levels, $U(G; d)$, aggregated from the mutation rate spectrum $U(R)$, the fitness model for membrane mutations, and the measured growth rates of target mutations. Arrows mark the maximum growth rate attainable by short-term evolution at a given population size. The low-growth component corresponds to unobservable low-resistance mutants. (c, d) Resulting normalized distributions of resistance effects and growth rates of mutant colonies in Luria-Delbrück assays, $P_{LD}(R; d)$ and $P_{LD}(G; d)$, at different drug levels. Filled squares indicate growth rate segments with probability > 0.04, dots mark observed mutants (as in Fig. 4a).

| Gene | Protein location | Pathway | Pathway function | Likely functional effect of mutation |
|---|---|---|---|---|
| *ispH* | Cytoplasm | Methylerythritol-phosphate pathway | Precursor for ubiquinol/ubiquinone biosynthesis pathway | Reduced ubiquinol / ubiquinone levels |
| *dxs* | Cytoplasm | Methylerythritol-phosphate pathway | Precursor for ubiquinol/ubiquinone biosynthesis pathway | Reduced ubiquinol / ubiquinone levels |
| *cydA-mngB* | Inner membrane | NADH – cytochrome oxidase electron transfer | Cytochrome bd-I ubiquinol oxidase subunit 1 | Reduced *cydA* expression |
| *cydB* | Inner membrane | NADH – cytochrome oxidase electron transfer | Cytochrome bd-I ubiquinol oxidase subunit 2 | Reduced ubiquinone levels |
| *ndh* | Inner membrane | NADH – cytochrome oxidase electron transfer | Quinone oxidoreductase enzyme | Impaired cytochrome oxidase |
| *ubiA* | Inner membrane | Ubiquinone biosynthesis pathway | Ubiquinone biosynthesis | Reduced ubiquinol / ubiquinone levels |
| *ubiI* | Cytoplasm | Ubiquinone biosynthesis pathway | Ubiquinone biosynthesis | Reduced ubiquinol / ubiquinone levels |
| *ubiH* | Cytoplasm | Ubiquinone biosynthesis pathway | Ubiquinone biosynthesis | Reduced ubiquinol / ubiquinone levels |
| *ubiK* | Cytoplasm / Inner membrane | Ubiquinone biosynthesis pathway | Ubiquinone biosynthesis | Reduced ubiquinol / ubiquinone levels |
| *atpG* | Inner membrane | ATP biosynthesis pathway | ATP synthase subunit involved in ATP synthesis | Impaired ATP synthesis |
| *hemC* | Cytoplasm | Heme biosynthesis pathway | Synthesis of cofactors for ubiquinol oxidases (*cydA, cydB*) | Reduced ubiquinol / ubiquinone levels |
| *arcB* | Cytoplasm / Inner membrane | NADH – cytochrome oxidase electron transfer, other pathways | Positive regulator of *cydA* expression (*50, 51*) | Reduced *cydA* expression |
| *selB* | Cytoplasm | Electron transfer | Translation factor, insertion of seleno-cysteine into membrane peptides | Modification of membrane-bound dehydrogenases |
| *ratA* | Cytoplasm | | Ribosome associated toxin / ubiquinone-binding protein | Reduced ubiquinone levels |

**Table S1: Membrane-associated resistance mutations: genes, pathways, and functions.** See reference (*52*) for a description of genes, encoded proteins, and their physiological role.

| Mutant number | $\frac{d_{\text{LD}}}{d_{50}^{\text{wt}}}$ | $W = \frac{\lambda_0}{\lambda_0^{\text{wt}}}$ | $\frac{d_*}{d_*^{\text{wt}}}$ | $\frac{\lambda_*}{\lambda_*^{\text{wt}}}$ |
|---|---|---|---|---|
| 1 | 0.9 | 0.87 (0.83, 0.91) | 4.31 (3.10, 4.86) | 1.79 (1.23, 2.14) |
| 2 | 0.9 | 0.82 (0.78, 0.86) | 3.42 (3.31, 3.42) | 1.68 (1.59, 1.74) |
| 3 | 0.9 | 0.93 (0.89, 0.96) | 2.70 (2.56, 2.88) | 1.38 (1.30, 1.42) |
| 4 | 0.9 | 0.63 (0.61, 0.66) | 0.74 (0.71, 0.80) | 0.14 (0.14, 0.16) |
| 5 | 0.9 | 0.88 (0.84, 0.92) | 1.52 (1.31, 1.57) | 1.47 (1.23, 1.51) |
| 6 | 0.9 | 1.00 (0.96, 1.04) | 0.43 (0.41, 0.44) | 0.29 (0.28, 0.31) |
| 7 | 0.9 | 0.22 (0.20, 0.25) | 2.78 (2.49, 3.02) | 0.34 (0.30, 0.45) |
| 8 | 1.8 | 0.72 (0.68, 0.77) | 3.15 (2.85, 3.20) | 0.99 (0.88, 1.02) |
| 9 | 1.8 | 0.87 (0.83, 0.92) | 3.73 (3.30, 3.93) | 2.62 (2.10, 3.10) |
| 10 | 1.8 | 0.53 (0.50, 0.55) | 0.06 (0.06, 0.07) | 0.01 (0.01, 0.01) |
| 11 | 1.8 | 0.78 (0.76, 0.82) | 2.97 (2.92, 3.06) | 1.00 (1.00, 1.05) |
| 12 | 1.8 | 0.63 (0.60, 0.67) | 3.40 (3.13, 3.46) | 0.70 (0.63, 0.72) |
| 13 | 1.8 | 0.73 (0.68, 0.76) | 3.36 (3.29, 3.65) | 0.87 (0.85, 0.92) |
| 14 | 1.8 | 0.92 (0.87, 0.97) | 4.19 (3.73, 4.33) | 1.84 (1.74, 2.37) |
| 15 | 1.8 | 0.88 (0.84, 0.93) | 2.12 (1.39, 2.19) | 1.39 (1.02, 1.48) |
| 16 | 3.7 | 0.68 (0.63, 0.74) | 6.20 (5.13, 6.86) | 2.06 (1.57, 2.62) |
| 17 | 3.7 | 0.49 (0.47, 0.53) | 6.84 (6.35, 7.14) | 1.02 (0.92, 1.09) |
| 18 | 3.7 | 0.60 (0.56, 0.64) | 6.84 (6.23, 7.83) | 0.98 (0.84, 1.04) |
| 19 | 3.7 | 0.68 (0.66, 0.71) | 4.00 (3.80, 4.11) | 0.72 (0.69, 0.76) |
| 20 | 3.7 | 0.51 (0.48, 0.54) | 6.83 (6.42, 7.08) | 1.04 (0.96, 1.12) |
| 21 | 3.7 | 0.58 (0.55, 0.61) | 4.97 (4.53, 5.01) | 0.69 (0.62, 0.72) |
| 22 | 3.7 | 0.58 (0.54, 0.61) | 4.98 (4.59, 5.18) | 0.78 (0.70, 0.80) |
| 23 | 3.7 | 0.57 (0.54, 0.61) | 6.20 (5.56, 6.58) | 0.78 (0.66, 0.84) |

| Mutant number | $\varepsilon = \frac{\gamma_{\text{in}}}{\gamma_{\text{in}}^{\text{wt}}}$ | $\frac{\gamma_{\text{out}}}{\gamma_{\text{out}}^{\text{wt}}}$ | $R = \frac{d_{50}}{d_{50}^{\text{wt}}}$ | $\frac{d_{\text{c}}}{d_{50}^{\text{wt}}}$ |
|---|---|---|---|---|
| 1 | 0.43 (0.38, 0.49) | 3.01 (1.58, 4.80) | 2.25 (2.09, 2.38) | 1.35 (1.14, 1.53) |
| 2 | 0.51 (0.47, 0.56) | 3.06 (2.55, 3.30) | 1.77 (1.67, 1.87) | 1.07 (0.97, 1.19) |
| 3 | 0.52 (0.49, 0.57) | 1.96 (1.71, 2.21) | 1.83 (1.73, 1.91) | 1.04 (0.94, 1.13) |
| 4 | 0.20 (0.19, 0.22) | 0.02 (0.02, 0.03) | 3.00 (2.83, 3.13) | 2.71 (2.52, 2.88) |
| 5 | 0.99 (0.90, 1.08) | 1.88 (1.52, 2.39) | 0.94 (0.88, 0.99) | 0.45 (0.39, 0.51) |
| 6 | 0.70 (0.66, 0.76) | 0.09 (0.08, 0.10) | 1.38 (1.31, 1.44) | 0.71 (0.64, 0.78) |
| 7 | 0.13 (0.11, 0.17) | 0.13 (0.09, 0.21) | 1.76 (1.53, 1.90) | 3.71 (3.15, 4.14) |
| 8 | 0.32 (0.30, 0.36) | 0.99 (0.79, 1.13) | 2.29 (2.16, 2.40) | 1.79 (1.62, 1.95) |
| 9 | 0.72 (0.59, 0.93) | 7.12 (4.56, 10.20) | 1.52 (1.32, 1.70) | 0.70 (0.47, 0.91) |
| 10 | 0.15 (0.13, 0.15) | 0.00 (0.00, 0.00) | 3.66 (3.46, 3.85) | 3.62 (3.37, 3.85) |
| 11 | 0.35 (0.33, 0.38) | 1.07 (0.98, 1.18) | 2.29 (2.17, 2.40) | 1.64 (1.52, 1.76) |
| 12 | 0.21 (0.20, 0.24) | 0.49 (0.41, 0.57) | 2.97 (2.79, 3.12) | 2.61 (2.40, 2.79) |
| 13 | 0.27 (0.24, 0.29) | 0.83 (0.72, 0.92) | 2.75 (2.59, 2.90) | 2.16 (1.98, 2.33) |
| 14 | 0.45 (0.42, 0.62) | 4.14 (3.07, 5.81) | 2.11 (1.78, 2.33) | 1.14 (0.85, 1.34) |
| 15 | 0.67 (0.62, 0.95) | 1.76 (1.09, 2.35) | 1.27 (0.99, 1.43) | 0.69 (0.47, 0.84) |
| 16 | 0.34 (0.27, 0.48) | 4.51 (2.54, 7.12) | 2.52 (2.11, 2.85) | 1.70 (1.16, 2.13) |
| 17 | 0.15 (0.14, 0.17) | 1.08 (0.87, 1.27) | 3.56 (3.34, 3.74) | 3.35 (3.07, 3.59) |
| 18 | 0.15 (0.11, 0.17) | 0.95 (0.72, 1.15) | 4.43 (3.84, 5.14) | 3.54 (3.07, 4.08) |
| 19 | 0.19 (0.17, 0.21) | 0.55 (0.48, 0.62) | 3.64 (3.40, 3.83) | 2.88 (2.67, 3.07) |
| 20 | 0.16 (0.14, 0.18) | 1.13 (0.93, 1.33) | 3.55 (3.33, 3.74) | 3.29 (3.02, 3.53) |
| 21 | 0.14 (0.13, 0.16) | 0.48 (0.40, 0.56) | 4.02 (3.74, 4.28) | 3.48 (3.21, 3.74) |
| 22 | 0.16 (0.14, 0.18) | 0.58 (0.49, 0.68) | 3.70 (3.49, 3.88) | 3.28 (3.04, 3.52) |
| 23 | 0.13 (0.11, 0.16) | 0.60 (0.46, 0.75) | 4.58 (3.99, 5.35) | 3.78 (3.31, 4.32) |

**Table S2: Growth and resistance data of membrane mutants.** (1) Mutant number. (2) Drug level of Luria-Delbrück assay, $d_{\text{LD}}/d_{50}^{\text{wt}}$. (3,4,5) Posterior average parameters of the membrane evolution model: resistance cost, $W = \lambda_0/\lambda_0^{\text{wt}}$; drug response parameters, $d_*/d_*^{\text{wt}}$, $\lambda_*/\lambda_*^{\text{wt}}$. (6,7) Membrane transport rates: uptake rate $\varepsilon = \gamma_{\text{in}}/\gamma_{\text{in}}^{\text{wt}}$; release rate, $\gamma_{\text{out}}/\gamma_{\text{out}}^{\text{wt}}$. (8) Resistance, $R = d_{50}/d_{50}^{\text{wt}}$. (9) Critical drug level, $d_c/d_{50}^{\text{wt}}$. All concentrations and rates are reported in units of the wild type parameters $d_{50}^{\text{wt}} = 8.66$ mg/L, $\lambda_0^{\text{wt}} = 2.02$, $d_{50}^{*\text{wt}} = 3.13$ mg/L, $\lambda_0^{\text{wt}} = 2/\text{h}$, $\lambda_*^{\text{wt}} = 0.37/\text{h}$. Measured growth rates and inferred growth inhibition curves are shown in Fig. S2. Inference procedures are detailed in section 3 of Methods.